%% file: tJ_prb.tex
\documentclass[reprint,aps,prl,superscriptaddress,amsmath]{revtex4-1}
\usepackage{graphicx}
\usepackage{amsmath}
\usepackage{amsfonts}
\usepackage{braket} 
\usepackage{color}
\RequirePackage[hyperindex,colorlinks,bookmarksnumbered, plainpages=true,pdfstartview=FitH]{hyperref}
\hypersetup{linkcolor=blue,urlcolor=blue,citecolor=blue}
\usepackage{hyperref}
\usepackage[export]{adjustbox}

\usepackage{color}
\usepackage{ulem}
\renewcommand{\emph}[1]{\textit{#1}}

\begin{document}
\title{iPEPS study of spin symmetry in the doped $t$-$J$ model}

\author{Jheng-Wei Li}
\affiliation{Arnold Sommerfeld Center for Theoretical Physics, Center for NanoScience,\looseness=-1\,  and Munich 
Center for \\ Quantum Science and Technology,\looseness=-2\, Ludwig-Maximilians-Universität München, 80333 Munich, Germany}
\author{Benedikt Bruognolo}
\affiliation{Arnold Sommerfeld Center for Theoretical Physics, Center for NanoScience,\looseness=-1\,  and Munich 
Center for \\ Quantum Science and Technology,\looseness=-2\, Ludwig-Maximilians-Universität München, 80333 Munich, Germany}
\affiliation{Max-Planck-Institut f\"ur Quantenoptik, Hans-Kopfermann-Strasse 1, D-85748 Garching, Germany}
\author{Andreas Weichselbaum} 
\affiliation{Arnold Sommerfeld Center for Theoretical Physics, Center for NanoScience,\looseness=-1\,  and Munich 
Center for \\ Quantum Science and Technology,\looseness=-2\, Ludwig-Maximilians-Universität München, 80333 Munich, Germany} 
\affiliation{
Department of Condensed Matter Physics and Materials Science, Brookhaven National Laboratory, Upton, NY 11973-5000, USA}
\author{Jan von Delft}
\affiliation{Arnold Sommerfeld Center for Theoretical Physics, Center for NanoScience,\looseness=-1\,  and Munich 
Center for \\ Quantum Science and Technology,\looseness=-2\, Ludwig-Maximilians-Universität München, 80333 Munich, Germany}

\begin{abstract}
We study the two-dimensional $t$-$J$ model on a square lattice using infinite projected entangled pair states (iPEPS).
At small doping, multiple orders, such as antiferromagnetic order, stripe order and superconducting order, are intertwined or compete with each other.
We demonstrate the role of spin symmetry at small doping by either imposing SU(2) spin symmetry or its U(1) subgroup in the iPEPS ansatz, thereby excluding or allowing spontaneous spin-symmetry breaking respectively in the thermodynamic limit.
From a detailed comparison of our simulations, we provide evidence that stripe order is pinned by long-range antiferromagnetic order.
We also find that SU(2) iPEPS, which enforces a spin-singlet state, yields a uniform charge distribution and favors d-wave singlet pairing.
\end{abstract}

\date{June 15, 2020}

\maketitle

\paragraph*{Introduction.}
The discovery of high-temperature superconductivity has triggered intense research on the properties of the one-band $t$-$J$ model on a square lattice, which has been argued to capture essential low-energy properties of cuprate materials \cite{Zhang_PRB_1988}. 
Despite many analytical and numerical works, full consensus regarding the competing low-energy states with different charge, spin and superconducting orders of the $t$-$J$ model has not yet been reached.
One category includes so-called stripe states, featuring spin-density waves and charge-density waves \cite{Poilblanc_PRB_1989, Zaanen_PRB_1989, Machida_1989, schulz1989domain, Emery_PRL_1990, Emery_PC_1993, Chetan_PRL_1997, White_PRL_1998, White_PRL1998b, White_PRB_1999, White_2000PRB, Eskes_PRB_1998, Pryadko_PRB_1999, White_PRB2000, White_PRB2002, Himeda_2002_PRL, CCP_PRB_2008, Yang_2009, Corboz_PRBR_2011, Sorella_PRL_2012, HW_PRB_2012, Corboz_PRL_2014}, where some of these states also exhibit coexisting $d$-wave superconducting order. 
Another potential candidate for the ground state of the hole-doped $t$-$J$ model is a superconducting state with uniform hole density \cite{Hellberg_PRL_1999, Raczkowski_PRB_2007,Chou_PRB_2008}.
Recently, Corboz \textit{et al.} \cite{Corboz_PRL_2014}, using infinite projected entangle pair states (iPEPS), demonstrated the energetically extremely close competition of the uniform state and the stripe state, even for the largest accessible numerical simulations.
Similar work on the Hubbard model also pointed towards a striped ground state \cite{Zheng_Science_2017, Huang_Nature_2018, Ido_PRB_2018, Darmawan_PRB_2018, Jiang2019, Ponsioen2019, Qin2019}.
Nevertheless, the underlying physical mechanism causing these intriguing ground-state properties remains illusive, and refined work in this direction is clearly necessary.

In this Rapid Communication, we focus on the so-called $\lambda5$ stripe state, featuring spin and charge modulations with a period of $\lambda=5$ lattice spacings, which has been previously shown to be energetically favorable near hole doping $\delta \sim 0.1$ at $J/t=0.4$ \cite[referred to as W5 stripe therein]{Corboz_PRL_2014}.
We use iPEPS (i) to study the evolution of $\lambda5$ stripe order from its optimal doping $\delta \sim 0.1$ into the spin and charge uniform phase; and (ii) to provide insight into the the relation between stripes and long-range antiferromagnetic (AF) order in the thermodynamic limit.

In particular, we show that by implementing either U(1) or SU(2) spin symmetry in the iPEPS ansatz, the relevance of long-range AF order can be directly examined.
Our analysis complements the finite-size scaling often used in density matrix renormalization group (DMRG) and Quantum Monte-Carlo (QMC) simulations, thereby addresses the question of ``the fate of the magnetic correlations in the 2D limit`` raised in Ref.~\onlinecite{Jiang_PRB_2018}.
Moreover, we show that the SU(2) iPEPS ansatz which, by construction, represents a spin singlet state, possesses d-wave singlet pairing order. 
Such SU(2) iPEPS can be interpreted as a generalized resonating valence bond (RVB) state \cite{Wang_PRL_2013, Poiblanc17, Poilblanc_PRB_2012, CJ_PRB_2018, Poilblanc_PRB2014}, and in this sense our finding of d-wave pairing for the SU(2) iPEPS is reminiscent of Anderson's original RVB proposal \cite{Anderson_PRL_1987, Kotliar_PRB_1988, Anderson2007}.

\bigskip
\paragraph*{Model and Methods.}
The $t$-$J$ Hamiltonian is given by
\begin{equation} 
  \label{eq:H_tJ}
  \hat{H} = -t \sum_{\braket{ij} \sigma} \big(\tilde{c}_{i \sigma}^{\dagger} \tilde{c}_{j \sigma} + \mathrm{H.c.} \big)
                 +J \sum_{\braket{ij}} (\hat{\bf{S}}_i \cdot \hat{\bf{S}}_j - \tfrac{1}{4} \hat{n}_i\hat{n}_j),
\end{equation}
with the projected fermionic operators $\tilde{c}_{i\sigma} = \hat{c}_{i\sigma} (1 - \hat{c}_{i\bar{\sigma}}^{\dagger} \hat{c}_{i\bar{\sigma}})$, spin operators $\hat{\bf{S}}_{i}$, spin label $\sigma \in \{ \uparrow, \downarrow \}$, and $\braket{ij}$ indexing all nearest-neighbor sites on a square lattice. 
We set $t=1$ as the unit of energy and use $J/t=0.4$, throughout.

We use iPEPS to obtain an approximate ground state for Eq.~(\ref{eq:H_tJ}).
The iPEPS ground state is a tensor network state consisting of a unit cell of rank-5 tensors, i.e., tensors with 5 indices or legs, repeated periodically on an infinite square lattice \cite{Verstraete04b, Verstraete_PRL_2006, Jordan2008, Kraus_PRA_2010, Corboz_PRB_2010, Bauer_PRB_2011, Corboz_PRL_2014, Singh_PRB2012, Liu_PRB2015, Poiblanc17, Schmoll_2018, Hubig18, Bruognolo19b, Schmoll_2020}.
Each rank-5 tensor has one physical index and four virtual indices (bonds) connecting to the four nearest-neighboring sites.
The accuracy of such a variational anstaz is guaranteed by the area law, and can be systematically improved by increasing the bond dimension $D$.

\begin{figure*} [thb]
 \centering
 \includegraphics[width=1\linewidth]{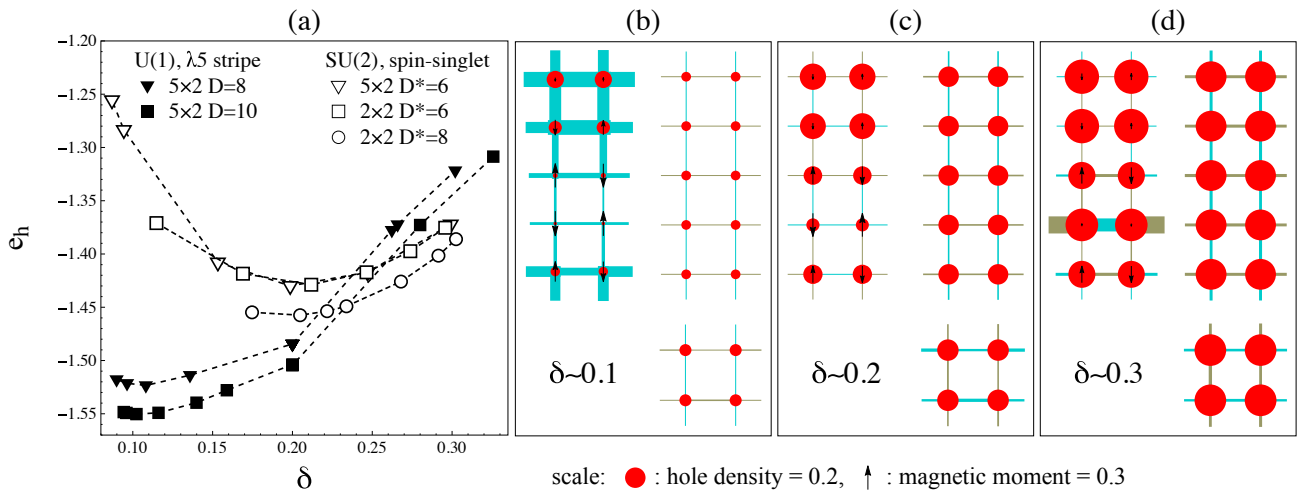}
 \caption{
    U(1) and SU(2) iPEPS results for the $t$-$J$ model at $J/t=0.4$.
    (a) The energy per hole, $e_{\rm h}$, as a function of hole doping $\delta$ for $5{\times}2$ and $2{\times}2$ unit cells.
    (b-d) Spin, hole and singlet pairing amplitude profiles at $\delta \sim 0.1$, $0.2$ and $0.3$, obtained using U(1) iPEPS yielding striped states(upper left), and using SU(2) iPEPS yielding uniform states($5\times2$: upper right, $2\times2$ lower right).
    Detail: the arrow height is proportional to the local magnetic moment, the radius of each red dot to the onsite hole density, and the bond thickness to the singlet pairing amplitude, with two colors indicating opposite signs.
}
\label{fig:tJ}
\end{figure*}

Using the QSpace tensor network library \cite{Wb12QS},
we can simply switch between exploiting either U(1) or SU(2) spin symmetries for our iPEPS implementation \cite{Bruognolo19b}.
This allows us to use sufficiently large bond dimensions to obtain accurate ground state wave functions.
With SU(2) iPEPS, we push the reduced bond dimension $D^{*}$ up to $8$, where $D^{*}$ is the number of kept SU(2) multiplets per virtual bond, which corresponds to a full bond dimension of $D\approx13$ states. 
To optimize the iPEPS wavefunctions via imaginary time evolution, we use full-update and fast full-update methods \cite {Jordan2008, Xie2009, Corboz_PRB_2010, Corboz_PRL_2014, Phien15}.
The contraction of the 2D infinite lattice is evaluated approximately by the corner transfer matrix (CTM) method \cite {Baxter_JSP_1978, Nishino_PLA_1996, Nishino_JPJ_1996, Orus09, Corboz_PRL_2014}, which generates so-called environment tensors with an environment bond dimension $\chi$.
For SU(2) iPEPS, the environment bond dimensions used here are $\chi^{*}=144$ $(\chi\approx 300)$ for $D^{*}=6$ $(D\approx11)$ or $\chi^{*}=128$ $(\chi\approx 270)$ for $D^{*}=8$ $(D\approx13)$.
For U(1) iPEPS, the environment bond dimensions are $\chi=256$ for $D=8$ or $\chi=200$ for $D=10$.

\bigskip
\paragraph*{Energetics.}
In Fig.~\ref{fig:tJ}(a), we show the energy per hole, $e_{\rm{h}}(\delta) \equiv (e_{\rm{s}} - e_{0})/\delta$, as a function of hole doping $\delta$, obtained from various iPEPS simulations (plots of $e_{\rm{s}}(\delta)$ vs. $\delta$ are shown in Ref.~\onlinecite{SI}, Fig.~S3).
Here $e_{\rm{s}}$ is the average ground state energy per site, and $e_{0} = -0.467775$ is the numerically exact value for the AF phase at zero doping taken from Ref.~\onlinecite{Sandvik_PRB_1997}.
Using U(1) iPEPS on a $5\times2$ unit cell, we find that the onset of $\lambda5$ stripe order occurs at $\delta \sim 0.1$, as previously reported \cite{Corboz_PRL_2014}.
Increasing the bond dimension from $D=8$ to $D=10$ improves the ground state energy consistently for every doping $\delta$ considered here. 
On the other hand, using SU(2) iPEPS ($D^{*}=6$), we obtain a spin-singlet state with no stripe feature on a $5\times2$ unit cell. 
Moreover, the ground state energy is almost independent of the shape of unit cells (cf.~$5\times2$ and $2\times2$ data).
We further improve the ground states using $D^{*}=8$ on the $2\times2$ unit cell.
Overall, for $\delta \lesssim 0.2$ in Fig.~\ref{fig:tJ}(a), we see that the U(1) $\lambda5$ stripe state yields a substantially lower ground state energy than the spin-singlet state, while the latter lies below the former for $\delta \gtrsim 0.25$.
From a technical perspective, our calculations show that for the non-symmetry-breaking phase favored at $\delta \gtrsim 0.25$, SU(2) iPEPS benefits from the full utilization of the spin-rotational symmetry, even though U(1) iPEPS has a larger number of variational parameters when $D>D^*$.

Next, we take a close look at each individual iPEPS for three values of doping.
The stripe states obtained using U(1) iPEPS, shown in the upper left parts of Figs.~\ref{fig:tJ}(b-d), exhibit modulation of charge and spin densities along the $y$ direction.
At $\delta \sim 0.1$, we find hole doping to be maximal along the top row, implying a site-centered stripe, in agreement with previous work \cite{Corboz_PRL_2014}.
Note that the spins in the two rows on either sides of the top row (row $2$ and $5$) are ordered antiferromagnetically (implying a so-called $\pi$ phase shift across the top row), thereby reducing the energy of transverse hole hopping along the domain wall \cite{White_PRL_1998, White_PRL1998b, White_PRB2002}.
At $\delta \sim 0.2$, we find hole doping to be maximal between two rows (the 1st. and 2nd.), implying a bond-centered stripe, as frequently observed in DMRG, DMET and QMC calculations \cite{White_PRL_1998, Zheng_Science_2017}.
Finally, at $\delta \sim 0.3$, the hole densities are roughly equal across all sites, with residual charge and spin modulation.
Overall, the stripe states we find here are in consensus with previous studies, which concluded that in the $t$-$J$ model stripe formation is predominantly driven by the competition between the kinetic energy and the exchange energy \cite{Zaanen_PRB_1989, Poilblanc_PRB_1989, Chetan_PRL_1997, White_2000PRB}. 
However, the same mechanism can also induce the pairing formation \cite{White_PRB_1999, white2000arxiv, Raczkowski_PRB_2007,Chou_PRB_2008}.
Therefore, it is a priori unclear that under what circumstances the system will favor stripe order or pairing at small doping.
To clarify this issue, we now turn to our SU(2) iPEPS results.

In contrast to the U(1) iPEPS results, switching on spin-rotational symmetry on the $5 \times 2$ unit cell by using SU(2) iPEPS suppresses the AF order and, hence, the spin modulation, as shown in the upper right parts of Figs.~\ref{fig:tJ}(b-d).
The resulting state no longer shows any spin stripes and instead has the same structure as the uniform state obtained on a $2\times2$ unit cell at similar doping (see the lower right parts of Figs.~\ref{fig:tJ}(b-d)). 
In addition, enforcing SU(2) symmetry also makes charge modulations completely disappear as well.
This observation suggests that in the $t$-$J$ model charge density waves (CDW) are strongly tied to spin stripes.

\begin{figure} [thb]
  \centering
  \includegraphics[width=0.95\linewidth,trim = 0.95in 0in 2in 0in, clip=true]{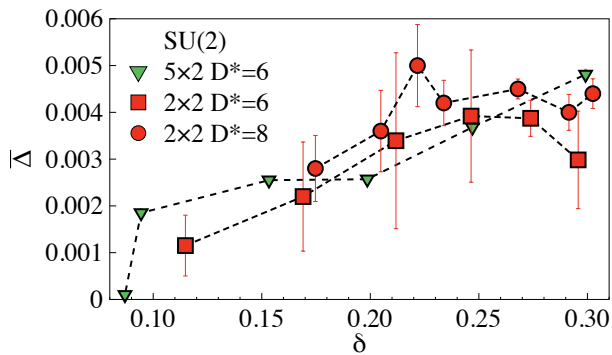}
  \caption{
    Averaged singlet pairing amplitude as a function of doping using SU(2) iPEPS.
    The error bar shows the mean absolute deviation of the pairing amplitudes among all bonds.
    For the $5\times2$ unit cell, the error bars are smaller than the symbols.
  }
  \label{fig:dwave}
\end{figure}

We have also examined d-wave superconducting order by computing the singlet paring amplitude, $\braket{\Delta_{ij}} = \frac{1}{\sqrt{2}}\braket{\tilde{c}_{i\uparrow}\tilde{c}_{j\downarrow}-\tilde{c}_{i\downarrow}\tilde{c}_{j\uparrow}}$.
For the U(1) iPEPS $\lambda5$ stripe states in Fig.~\ref{fig:tJ}(b-d), we cannot directly identify a d-wave pairing character, in contrast to Refs.~[\onlinecite{Corboz_PRL_2014},~\onlinecite{Zheng_Science_2017}], which found opposite signs for the amplitude of the bonds along the $x$- and $y$-axis.
However, a word of caution is necessary in reading this result when the ground state spontaneously breaks SU(2) spin symmetry,
because even a trivial term, such as $\braket{\tilde{c}_{i\uparrow}\tilde{c}_{j\downarrow}}$, could yield a non-zero contribution to $\braket{\Delta_{ij}}$.
For a more rigorous diagnosis, one should explicitly study the pair correlation function \cite{Dagotto_PRB_1992, Dagotto_PRB_1994, Jiang_PRB_2018, CC_PRB_2018}, which goes beyond the scope of this work.
Hence, our results do not exclude the possibility that stripes and d-wave superconducting order could coexist.

On the other hand, the SU(2) iPEPS is a spin-singlet state, by construction.
It takes into account short-range spin correlations but excludes long-range AF order, which breaks spin-rotational symmetry in the thermodynamic limit (see Supplementary Information for details).
This rules out the aforementioned ambiguity, and the singlet pairing amplitude becomes a robust measure.
As shown in Figs.~\ref{fig:tJ}(b-d), a d-wave pattern appears on both $5\times2$ and $2\times2$ unit cells.
Fig.~\ref{fig:dwave} shows the averaged singlet pairing amplitude, $\overline{\Delta} = \frac{1}{N}\sum_{\braket{ij}}f({\bf{r}}_{ij})\braket{\Delta_{ij}}$
as a function of doping, where $N$ is the number of sites in the unit cells, ${\bf{r}}_{ij} \equiv {\bf{r}}_{j} - {\bf{r}}_{i}$,
and $f({\bf{r}})$ is a d-wave form factor, which takes the values $f(\pm\hat{{\bf{y}}})=-1$ and $f(\pm{\hat{\bf{x}}})=1$, respectively.
The error bar indicates the mean absolute deviation of the pairing amplitudes among all bonds.
In the $2\times2$ case, the pronounced deviation is mostly attributed to the difference in pairing amplitudes along the $x$ and $y$ directions.
A similar phenomenon has also been observed in a recent large-scale DMRG calculation \cite{Jiang_PRB_2018}, and an almost equal mixture between d-wave and s-wave singlet paring amplitude has been suggested.
Upon increasing the bond dimension $D^*$ from $6$ to $8$, the d-wave pairing order increases.
This is different from the previous analysis of charge uniform states using U(1) iPEPS, where pairing is suppressed with increasing $D$ \cite{Corboz_PRL_2014}.
Furthermore, the $5\times2$ case also shows a rather uniform d-wave pattern.
All in all, our SU(2) iPEPS results provide direct evidence that the doped $t$-$J$ model exhibits d-wave superconductivity in the thermodynamic limit.

\bigskip
\paragraph*{Influence of stripes on antiferromagnetic order.}
In the previous section we have shown that stripes can be stabilized as ground states using the U(1) iPEPS at doping $0.1 \lesssim \delta \lesssim0.2$ on a $5\times2$ unit cell.
By contrast, the SU(2) iPEPS shows no signature of any spatial modulations of spin and charge density.
This suggests that the stripes and the AF order are intimately related.
While such a viewpoint has been discussed extensively both theoretically and experimentally since the discovery of the so-called $\frac{1}{8}$ anomaly \cite{Tranquada_NATURE_1995, Kivelson_RMP_2003, Vojta_2009AIP, Robinson2019}, a direct understanding of how AF order coexists with stripes is still lacking.

To address this, we have computed the staggered spin-spin correlation functions for the ground state,
  \begin{align}
    C(i) = \tfrac{(-1)^{x+y}}{\frac{3}{4}(1-\delta)N}
    \sum_{j\in \text{unit cell}}
    \bigl\langle \hat{\bf{S}}_{j+i} \cdot \hat{\bf{S}}_j \bigr\rangle ,
  \end{align}
with $i=(x,y)$. The prefactor normalizes the same-site correlator
to unity, $C(0)=1$,
given $(1-\delta)N$ spins per unit cell.
This facilitates the comparison of different unit cells and doping.
In the following, we analyze $C(i)$ along the long ($y$) and short ($x$) directions of the unit cell.

\begin{figure} [tb!]
  \centering
  \includegraphics[width=1\linewidth,trim = 2in 2.5in 2.5in 2in, clip=true]{Figures/C_5x2.pdf}
  \caption{
  Normalized staggered spin-spin correlation functions, computed on a $5\times2$ unit cell along the long ($y$) direction (left column) and the short ($x$) direction (right column), using U(1) iPEPS (top row) and SU(2) iPEPS (lower two rows), on linear and semilogarithmic scales, respectively.
  [Filled symbols indicate the variational state (U(1) or SU(2)) having the lower energy for a given $\delta$.] 
  }
  \label{fig:Correlation5x2}
\  \\[2ex]
  \includegraphics[width=1\linewidth,trim = 2in 1in 2in 6in, clip=true]{Figures/C_Wx2.pdf}
  \caption{
    Comparison of normalized staggered spin-spin correlation functions
    for $\delta\simeq0.2$ using U(1) iPEPS at $D=8$ along the long ($y$) direction and the short ($x$) direction on $L\times 2$ unit cells for $L=2,3,4,5$ in (a-d), respectively.
    Inset of (b) is a $\lambda4$ stripe state obtained from different initialization.
  }
  \label{fig:CorrelationWx2}
\end{figure}

First, we study the staggered spin-spin correlations on a $5\times2$ unit cell at doping $\delta = 0.1, 0.2$ and $0.3$, using U(1) iPEPS.
In Fig.~\ref{fig:Correlation5x2}(a), we can clearly identify $\lambda5$ stripe order at $\delta = 0.1$ and $0.2$, with staggered spin-spin correlations oscillating around zero, reflecting the pattern already seen in the left panels of Figs.~\ref{fig:tJ}(b,c).
The staggered magnetic order undergoes a phase shift of $\pi$ across the length of the $5\times2$ unit cell, resulting in a period of $\lambda_m=10$. 
At doping $\delta = 0.3$, the correlations decay much more rapidly, with weak residual oscillations remaining at large distances.
Given its higher variational energy compared to its SU(2) counterpart, this reflects the numerical inefficiency of using a broken-symmetry ansatz to simulate a spin singlet when many low-energy states are nearly degenerate.

By contrast, Fig.~\ref{fig:Correlation5x2}(b) shows that the correlations along the ``short'' direction decrease with doping, but remain positive at large distances, indicating long-range AF order, i.e., $C(|i|\rightarrow\infty)\neq0$, yet attenuated with increasing $\delta$.
Therefore, Figs.~\ref{fig:Correlation5x2}(a,b) suggest that stripes along the long direction go hand in hand with long-range AF order along the short direction.

To further elucidate this point, we turn our attention to the SU(2) iPEPS.
Again, we have computed the staggered spin-spin correlations on a $5\times2$ unit cell using SU(2) iPEPS.
In Figs.~\ref{fig:Correlation5x2}(c,d), the correlations along the long and short directions are nearly identical and rapidly decay to zero, showing no sign of either stripes or the long-range AF order. 
Note that for SU(2) iPEPS, the instability of a given state towards AF order can be detected by the increase of correlation length when increasing $\chi^*$. (We illustrate this for the Heisenberg model in  Ref.~\cite{SI}~Sec.~SI). 
However, this tendency is not observed at $\delta=0.1$ (see Figs.~\ref{fig:Correlation5x2}(e,f)).
In short, we conclude that stripes only emerge in the presence of long-range AF order.

To strengthen our previous statement, we further consider $L\times2$ unit cells with $L=5,4,3,2$ at $\delta \sim 0.2$ using U(1) iPEPS ($D=8$).
Those could host spin stripes of periods $\lambda=L$, or an AF ordered state for $L=2$.
A previous iPEPS study has shown a very close competition between a $\lambda5$ stripe state and an AF state with uniform charge distribution ($L=2$) at $\delta \sim 0.1$ \cite{Corboz_PRL_2014}. 
As the preferred stripe period decreases with increasing doping \cite{Darmawan_PRB_2018}, here we focus on stripe periods $\lambda \leq 5$ at $\delta \sim 0.2$.
For a $2\times2$ unit cell [Fig.~\ref{fig:CorrelationWx2}(a)], the spin-spin correlations along both long and short directions quickly decay to nearly zero, showing that AF order is weak at $\delta\sim0.2$ if a charge-uniform state is assumed.
The same charge-uniform state is also favored for a $4\times2$ unit cell: we obtain this by initializing the $4\times2$ unit cell of a full-update optimization using two copies of the $2\times2$ unit cell of panel (a), which yields a slightly lower energy than a $\lambda4$ stripe sate [inset of (b)] initialized from simple-update results.
By contrast, $3\times2$ and $5\times2$ unit cells show a clear stripe feature along the long direction, together with non-zero long-range AF order along the short one [Figs.~\ref{fig:CorrelationWx2}(c,d)], and slightly lower ground state energy than those of $2\times2$ and $4\times2$.
However, the bond dimension $D$ used here is not large enough to conclusively resolve the close competition between the different states.
Overall, by plotting the correlations along both short and long directions in the same panel, we see that the amplitude of the stripe modulation is the same as that of attenuated long-rage AF correlations.
This further confirms that the stripes and the long-range AF order are indeed tied to each other at finite doping.

\bigskip
\paragraph*{Summary.}
We have studied the doped $t$-$J$ model with $J/t=0.4$ using U(1) and SU(2) iPEPS.
For doping $0.1 \lesssim \delta \lesssim0.2$, $\lambda$5 striped charge and spin order with U(1) symmetry is energetically favorable compared to a spin-singlet state with SU(2) symmetry.
By contrast, for $\delta \gtrsim 0.25$, the latter is favored.
By studying the spin-spin correlations, we find a close link between stripe order and long-range AF order.
At small doping, the U(1) iPEPS shows that spin stripes emerge along one spatial direction, while attenuated long-range AF order persists along the other spatial direction.
Upon increasing doping, the strength of stripe order decreases hand in hand with long-range AF order.
By contrast, the SU(2) iPEPS, which does not break spin rotational symmetry, excludes long-range AF order and, hence, stripe formation, but yields d-wave superconducting order at finite doping.
Our study demonstrates the utility and importance of being able to turn on and off the SU(2) spin-rotational symmetry at will -- it gives direct insights into the interplay between regimes with spontaneously broken symmetries and where SU(2) invariance remains intact.

\bigskip
\paragraph*{Acknowledgments.}
The Deutsche Forschungsgemeinschaft supported BB, JWL and JvD through the Excellence Cluster ``Nanosystems Initiative Munich'' and the Excellence Strategy-EXC-2111-390814868, and AW through WE4819/2-1.  
AW was also supported by DOE-DESC0012704.
\input{tJ_prb.bbl}

\appendix
\clearpage

\setcounter{page}{1}
\thispagestyle{empty}

\onecolumngrid

\begin{center}
\vspace{0.1cm}
{\bfseries\large Supplementary Material \--- iPEPS study of spin symmetry in the doped $t$-$J$ model}\\
\vspace{0.4cm}
Jheng-Wei Li,$^1$
Benedikt Bruognolo,$^{1,2}$
Andreas Weichselbaum,$^{1,3}$
and Jan von Delft,$^1$
\\
\vspace{0.1cm}
{\it 
$^1$Arnold Sommerfeld Center for Theoretical Physics, 
Center for NanoScience,\looseness=-1\,  and 
Munich Center for \\ Quantum Science and Technology,\looseness=-2\, 
Ludwig-Maximilians-Universit\"at M\"unchen, 80333 Munich, Germany\\
$^2$Max-Planck-Institut f\"ur Quantenoptik, Hans-Kopfermann-Strasse 1, D-85748 Garching, Germany\\
$^3$Department of Condensed Matter Physics and Materials Science,\\ Brookhaven National Laboratory, Upton, NY 11973-5000, USA
} \\
\date{June 15, 2020}
\vspace{0.6cm}
\end{center}

\twocolumngrid
In this Supplementary Material, we include discussions of the 2D Heisenberg model, phase separation, and additional details of our simulations.

\section{Correlations of the 2D Heisenberg model}
In the main text, we point out the difference between U(1) iPEPS and SU(2) iPEPS in the doped $t$-$J$ model.
Here, we further illustrate that for $\delta=0$.
It is known that, at zero doping, the $t$-$J$ model reduces to the antiferromagnetic Heisenberg model, and the ground state exhibits spontaneous symmetry breaking.
At this critical point, we show that the U(1) iPEPS has finite AF magnetization, and hence rigid long-range order.
In contrast, the SU(2) iPEPS can not have long-range order in a given ground state, even though the paramagnetic phase is unstable and the criticality can be inferred by the slow decay of the staggered spin-spin correlation function.

In Tabel~\ref{tabel:Heisenberg} we summarize our results for the 2D AF Heisenberg model obtained from U(1)-symmetric and the SU(2)-symmetric iPEPS calculations, using $J=1$ as unit of energy.
Our U(1) iPEPS variational energy per site, $e_0=-0.6693$, agrees well with the best estimate from QMC \cite{Sandvik_PRB_1997}, namely $-0.6694$.
The SU(2) simulations, by construction, represent a symmetrized state and hence cannot gain energy from spontaneously symmetry breaking. 
Therefore they yield a slightly higher energy, consistent with previous works \cite{Hubig18, Poiblanc17}.

\begin{table}[b!]
  \caption{
  Comparison of the ground state energy per site $e_\mathrm{0}$, and the square of the local magnetization $m_{\text{stag}}^{2}$, for the 2D Heisenberg model between U(1) and SU(2) iPEPS results.
  We report values obtained at the largest possible CTM environment in the measurement, for U(1) $D=4$ and $8$, $\chi=512$ and, for SU(2) $D^{*}=4$, $\chi^{*}=128$.}
  \label{tabel:Heisenberg}
  \begin{ruledtabular}
     \begin{tabular}{llll}
     &  $e_0$ & $m_{\text{stag}}^{2}$
     \\\hline \\[-2mm]
     \ \,\,$\text{U(1)}_{D=4}$ &  $-0.6686$ & $0.1271$  \\
     \ \,\,$\text{U(1)}_{D=8}$ &  $-0.6693$ & $0.1141$  \\
     $\text{SU(2)}_{D^{*}=4\ (D=12)}$ & $-0.6686$ & --- 
     \\
     \end{tabular}
  \end{ruledtabular}
\end{table}

\begin{figure}
  \centering
  \includegraphics[width=1\linewidth,trim = 2in 1in 2in 0in, clip=true]{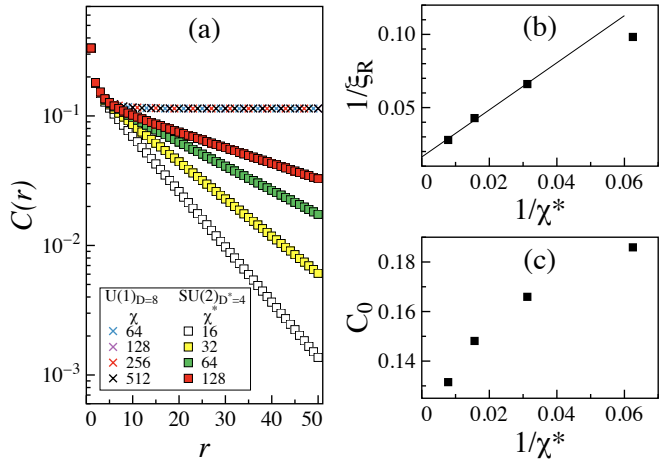} 
  \caption{
  (a) Staggered spin-spin correlation functions
  (b) The inverse of correlation length $1/\xi_{\text{R}}$ extracted from the asymptotic large-distance behaviors $C(r)\sim C_0 e^{-r/\xi_{\rm{R}}}$vs $\chi^*$ for SU(2) iPEPS
  (c) $C_0$ vs $1/\chi^*$ for SU(2) iPEPS
}
  \label{fig:Heisenberg}
\end{figure}

Next, we study how symmetry affects the real-space spin-spin correlations. 
Fig.~\ref{fig:Heisenberg}(a) shows the staggered spin-spin correlations for both U(1) and SU(2) symmetry, for several values of $\chi$, on a semilogarithmic scale.
For the U(1) iPEPS, the correlations quickly saturate to a nonzero value with increasing distance, resulting in a finite magnetization $\lim\limits_{r\rightarrow\infty} C(r) = m_{\text{stag}}^{2}$, and remain almost the same upon increasing $\chi$ from 64 to 512.
This suggests that these CTM simulations are already well converged with respect to $\chi$  for the given finite value of $D$.
The subleading spin-spin correlations on top of this AF background can be extracted by computing 
  $\delta C_{ij} = \langle \hat{\bf S}_i \cdot \hat{\bf S}_j \rangle - \langle \hat{\bf S}_i\rangle \cdot \langle \hat{\bf S}_j \rangle$ (not shown), which decay exponentially with $r\geq5$ lattice spacings.

For the case of SU(2) iPEPS, on the other hand, it holds by construction that $\langle \hat{\bf S}_j \rangle=0$.
Therefore, $C(r)$ itself decays exponentially at sufficiently large distance $r$.
In either case, $\delta C$ for U(1) or $C$ for SU(2) here, this implies finite gaps induced by the finite bond dimensions $D$ and the finite environment dimensions $\chi$.
However, for the case of SU(2), the slope of the exponential decay has a strong dependence on $\chi$.
The correlation length increases with increasing $\chi$, and no sign of saturation is found up to $\chi^{*}=128$ ($\chi\approx500$), as expected for a critical state with no gap (see Fig.~\ref{fig:Heisenberg}(b,c)).

Recently, it has also been pointed out that SU(2) iPEPS is capable of capturing the critical phase in the study of the $J_1$-$J_2$ Heisenberg model \cite{Poiblanc17}.
However, we notice a subtle difference regarding the issue of symmetry-breaking between our simulations and theirs.
In their SU(2) iPEPS simulations for $D^{*}=4$, the CTM approximation at finite $\chi$ induces symmetry breaking, and the resulting spurious staggered magnetization only vanishes when $\chi\rightarrow\infty$.
In contrast, no such observation of symmetry-breaking appears in our SU(2) iPEPS calculations.
We suspect the difference comes from the different setup of the PEPS ansatz.
In Ref.~\onlinecite{Poiblanc17}, a single-site PEPS with $C_{4v}$ rotational symmetry is assumed, and the virtual multiplets occuring in the ansatz are manually preselected. 
For $D^{*}=4$, they are fixed to $\{0,(\frac{1}{2})^3\}$ (here the superscript specifies the multiplicity, i.e., the number of multiplets in a given symmetry sector, resulting in $D=1+3\cdot2 = 7$ states total).

\begin{figure}
  \centering
  \includegraphics[width=0.2\textwidth]{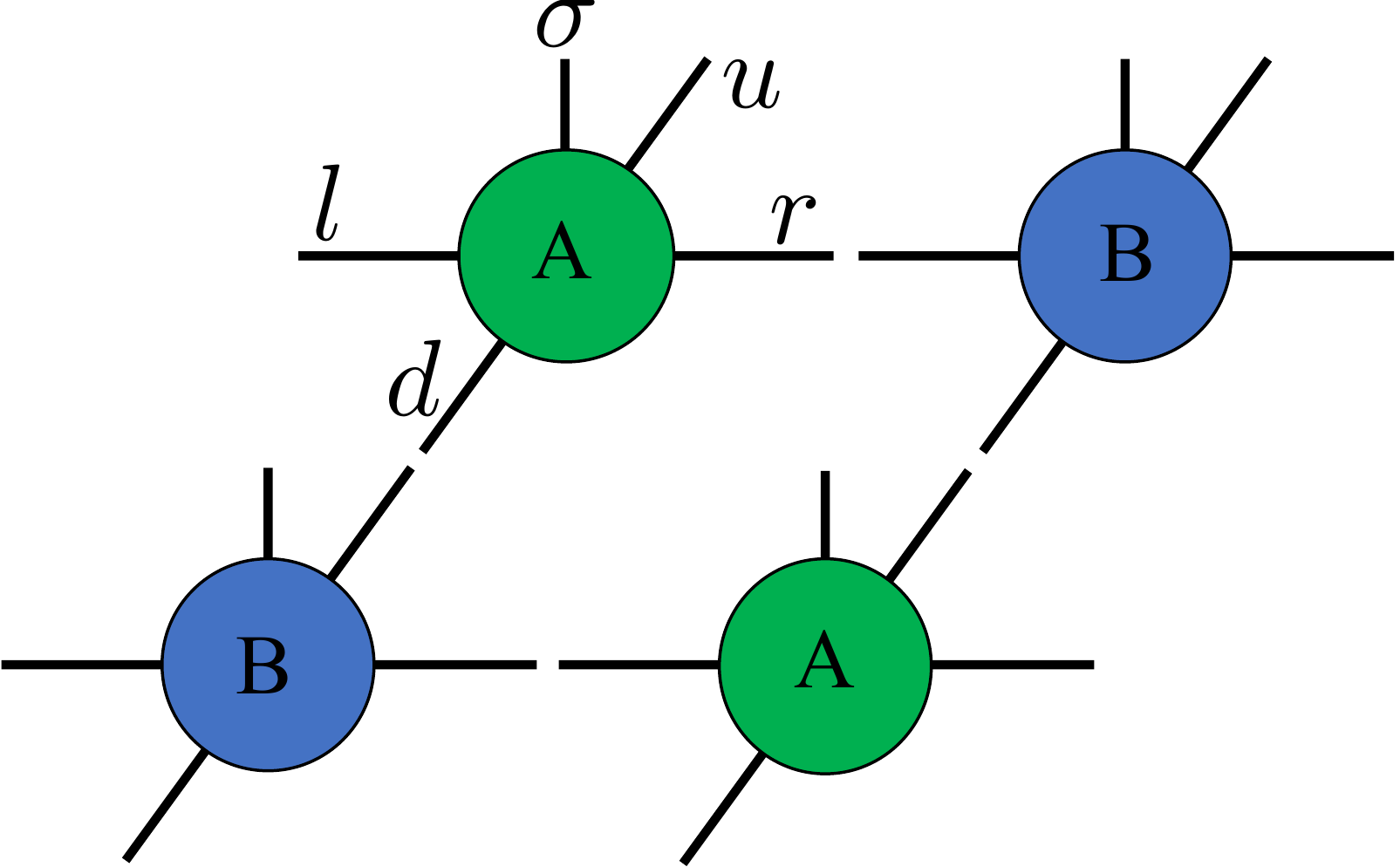}
  \caption{
     A pictorial representation of an iPEPS on a $2\times2$ lattice with an $ABBA$ tiling pattern. $\sigma, l, u, d, r$ represents the physical index and four virtual indices at left, up, down and right directions, respectively. 
  }
  \label{fig:ABBA} \vspace{-5mm}
\end{figure}

Instead, we use a $2\times2$ unit cell with $4$ independent tensors, and we allow the quantum numbers at virtual legs to fluctuate during optimization.
In our setup, the converged $2\times2$ PEPS shows an $ABBA$ tiling pattern (see Fig.~\ref{fig:ABBA}). 
The variationally selected virtual space for $D^{*}=4$ leads to the multiplets $\{0, 1^2, 2\}$ (i.e., $D=1+2\cdot3 + 5 = 12$) for three virtual legs, and $\{(\frac{1}{2})^2,(\frac{3}{2})^2 \}$ (i.e., $D=2\cdot2 + 2\cdot4 = 12$) for the fourth one.
This seems to break the translational and rotational symmetry for a single site.
However, all bonds have the same total bond dimension of $D=12$ states.
(Note that, if the local state space has half-integer spin, it is not possible to have all-integer or all half-integer spins on all virtual bonds.
Of course, the iPEPS tensors could be symmetrized to have a linear combination of integer and half-integer spins on each of the virtual bonds.
But \textit{a priori}, having the above configuration with half-integer spins on only one virtual leg in the SU(2) iPEPS does not necessarily imply that the state is physically anisotropic.)

In short, we have addressed the key difference between U(1) and SU(2) iPEPS for the 2D AF Heisenberg model.
U(1) iPEPS ($D=8$) exhibits long-range AF order, breaking spin rotational symmetry, and the quantum fluctuations are short-ranged.
In contrast, SU(2) iPEPS ($D^{*}=4$) is critical: at any finite $\chi$, there is no long-range AF order, but the AF spin fluctuations are strong and slowly decaying.
Taking $\chi\rightarrow\infty$, we find a diverging correlation length $\xi_{\text{R}}$, as expected for quantum criticality.

\vspace{-1mm}

\section{Comments on phase separation}

\vspace{-1mm}

The energy per hole $e_h(\delta)$ has often been used to detect the stability of a given phase at small doping relative to the AF phase at the zero doping. \cite{Emery_PRL_1990, White_PRB2000, HW_PRB_2012, Corboz_PRL_2014}.
It has been argued that if $e_{h}(\delta)=(e_{s}(\delta)-e_0)/\delta$ has a minimum at $\delta_c$, the region $\delta \lesssim \delta_c$ is unstable, with a tendency towards phase separation.

\begin{figure}
  \centering
  \includegraphics[width=1\linewidth,trim = 8in 6in 6in 3in, clip=true]{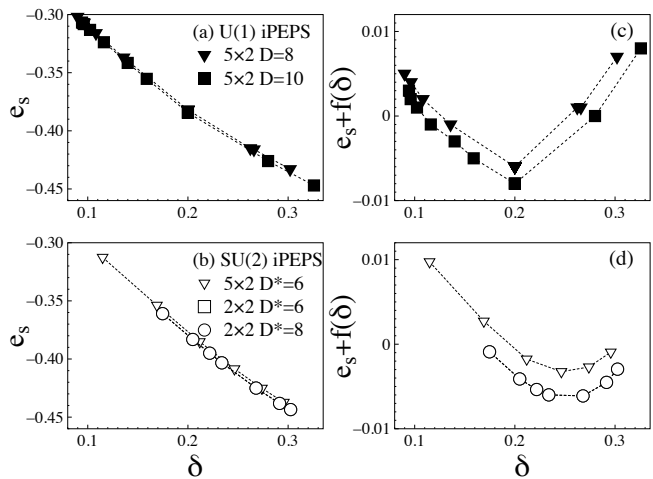}
  \caption{
     (a,b) Energy per site, $e_s$, as a function of hole doping $\delta$ for U(1) and SU(2) iPEPS.
     (c,d) Same data, but subtracting the linear slope by adding $f(\delta) = 0.63\,\delta+0.25$ 
      to improve the visibility of the curvature.
  }
\vspace{-6mm}
  \label{fig:tJ_Es}
\end{figure}

Here we point out that this argument is only valid at the dilute limit, i.e, $\delta \ll 1$.
For a system to phase separate, \textit{the energy per site, $e_{s}$, must have a negative curvature, $e_{s}''(\delta)<0$}, or, equivalently, $e_{s}''= (\delta \cdot e_{h} +e_0)'' = 2e_h'+\delta e_h''<0$.
Now, for sufficient small doping and assuming that $e_h''$ is bounded, the second term $\delta e_{h}''$ becomes negligible, so that $e_{s}'' \approx 2e_h'$.
In this case, phase separation ($e_{s}''<0$) implies $e_{h}'<0$, so that a minimum in $e_h$ at $\delta_c$ (i.e., a sign change in $e_{h}'$ from $>0$ to $<0$ indeed implies phase separation.
However, this is no longer the case if $\delta$ becomes sizable.
Indeed, consider Fig.~\ref{fig:tJ_Es}, where we have replotted the data for $e_h(\delta)$ vs. $\delta$ from Fig.~1(a) of the main text, but now showing $e_s(\delta)$ vs. $\delta$.
In the simulated doping range, we find $e_{s}''\geq0$, i.e., no indication of phase separation, even though the SU(2) results for $e_{\rm{h}}$ shows a clear minimum near $\delta \sim 0.2$.

\vspace{-1mm}

\section{Further symmetry and filling related technicalities}

\vspace{-1mm}

In this section, we document the  details of our iPEPS simulations for the doped $t$-$J$ model with $J/t=0.4$.
In our setup, the total particle number is not conserved, and $\mathbb{Z}_{2}$ parity symmetry is used in the charge sector.
The average number of holes is controlled by the chemical potential $\mu$.
In order to ensure that at $\mu=0$ the system half-filled, we add an additional term, $\frac{2t^{2}}{J}\sum_{i}(\hat{n}_{i}-\frac{1}{2})$, to the $t$-$J$ model \cite{Dagotto_PRB_1992}.
This term is similar to the one used in the Hubbard model, $\frac{U}{2}\sum_{i}(\hat{n}_{i}-\frac{1}{2})$, to make the Coulomb interaction particle-hole symmetric.
The spin sector, as discussed in the main text, has either U(1) or SU(2) symmetry.

\onecolumngrid
\section{Further figures}
For reference, Tables~\ref{table:U(1)}-\ref{table:SU(2)} depict detailed numerical values for the  iPEPS states shown in Fig.~1(b-d), and several related states. 
\begin{table*}[htb!]  
  \centering
  \caption{
     U(1) iPEPS with $D=8$ and $D=10$ at doping $\delta\sim 0.1$, $0.2$ and $0.3$. 
     The left (right) column of numbers give hole densities (local moments), averaged (in terms of absolute value) over the two sites in a row.
  } \vspace{3mm}
  \label{tabel:XXX}
     \begin{tabular*}{.65\textwidth}{|c|c|c|}
     \hline
     $ \delta$ & $D=8$ & $D=10$ \\
     \hline
     $0.1$ & 
     \includegraphics[valign=c,width=0.3\textwidth]{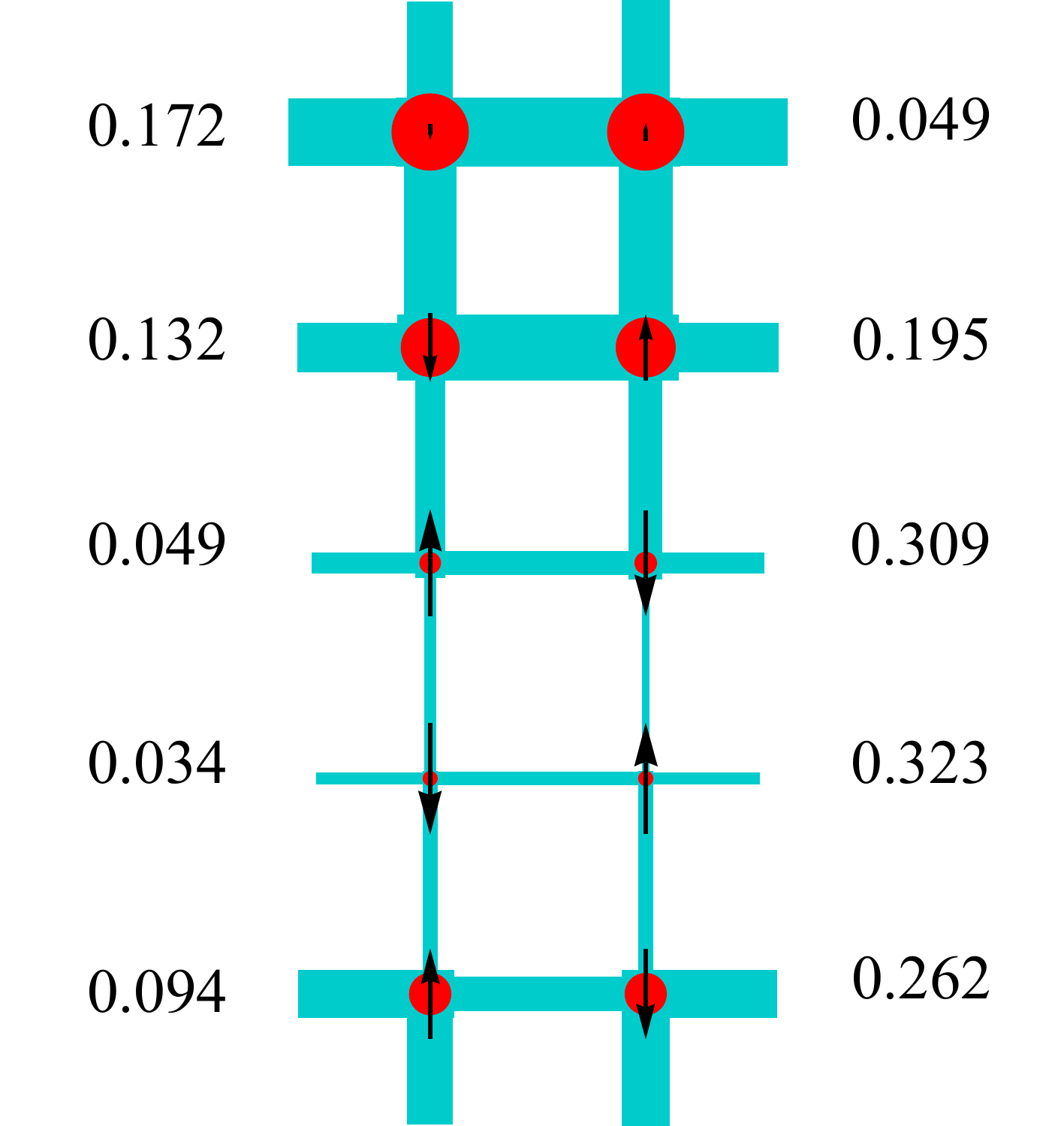}   & 
     \includegraphics[valign=c,width=0.3\textwidth]{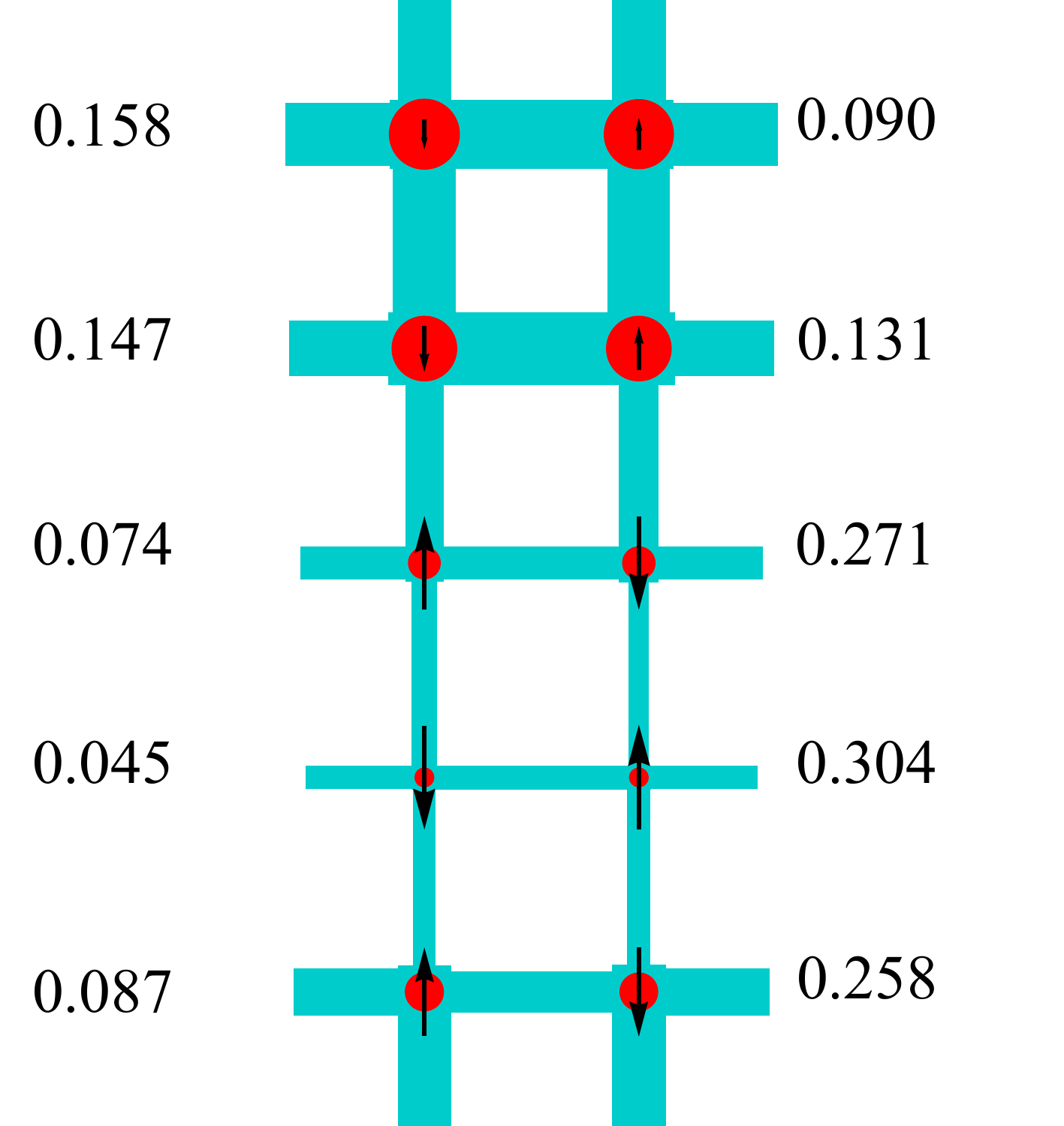}\\
     \hline
     $0.2$ & 
     \includegraphics[valign=c,width=0.3\textwidth]{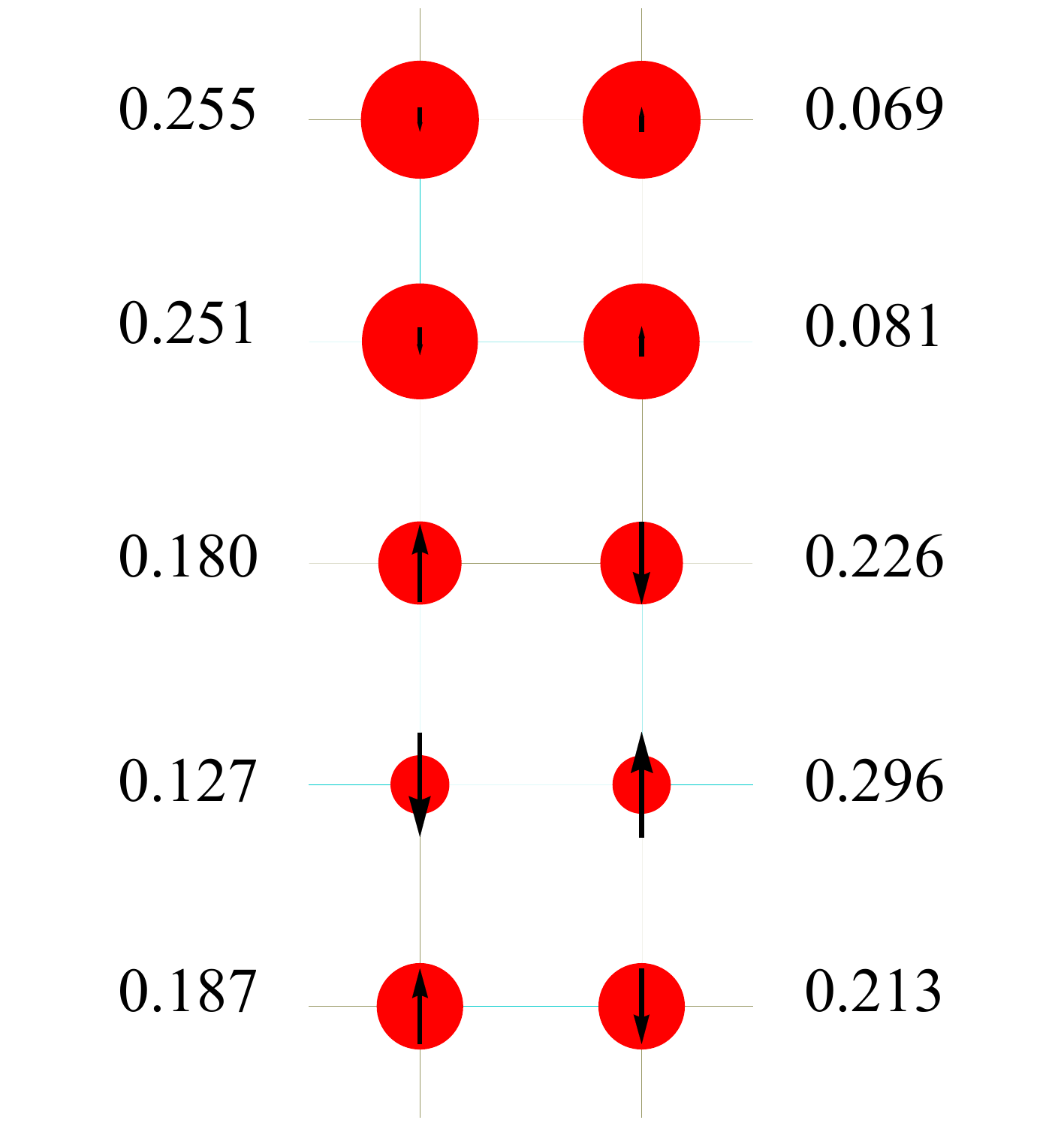}    & 
     \includegraphics[valign=c,width=0.3\textwidth]{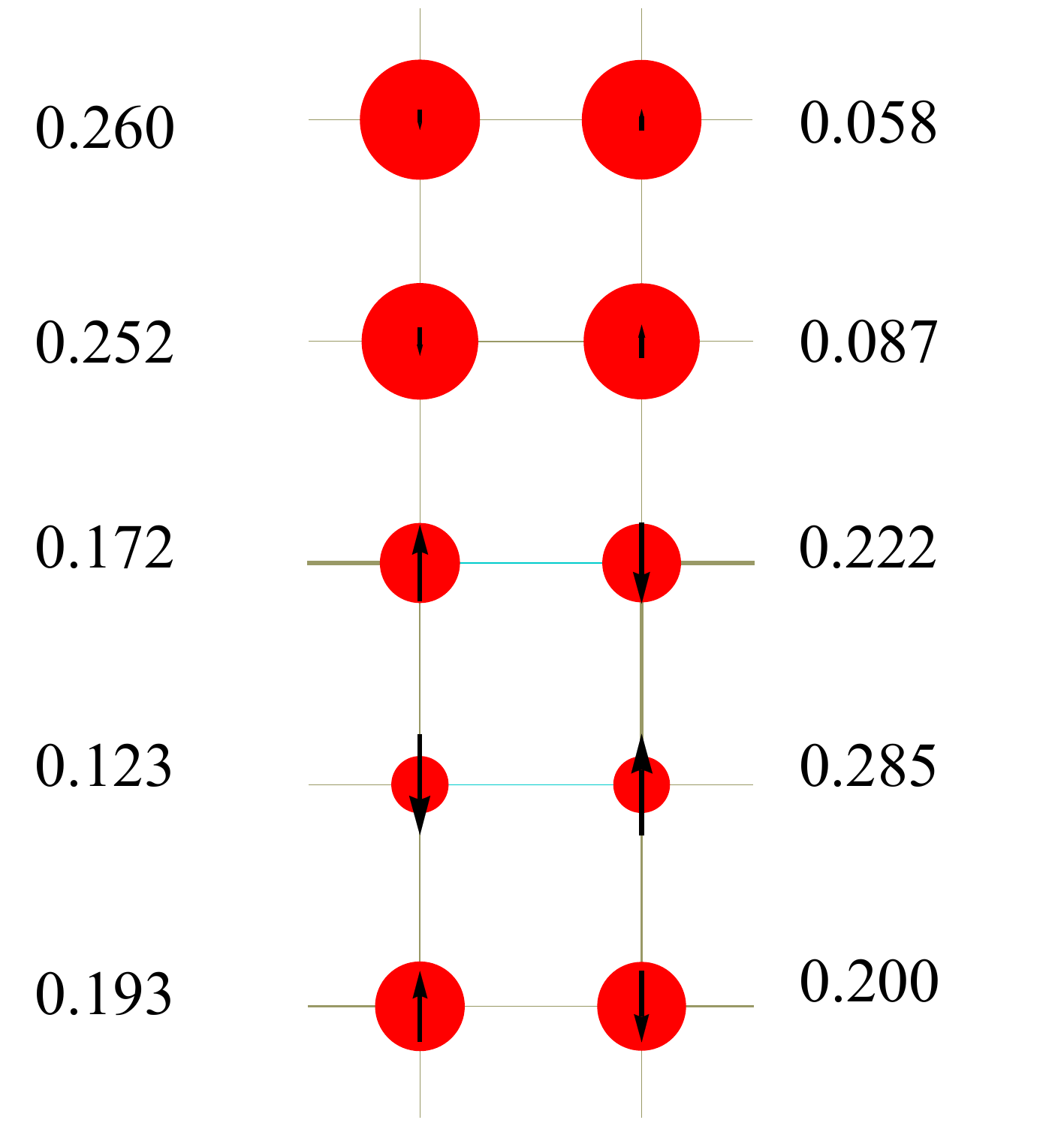} \\
     \hline
     $0.3$ & 
     \includegraphics[valign=c,width=0.3\textwidth]{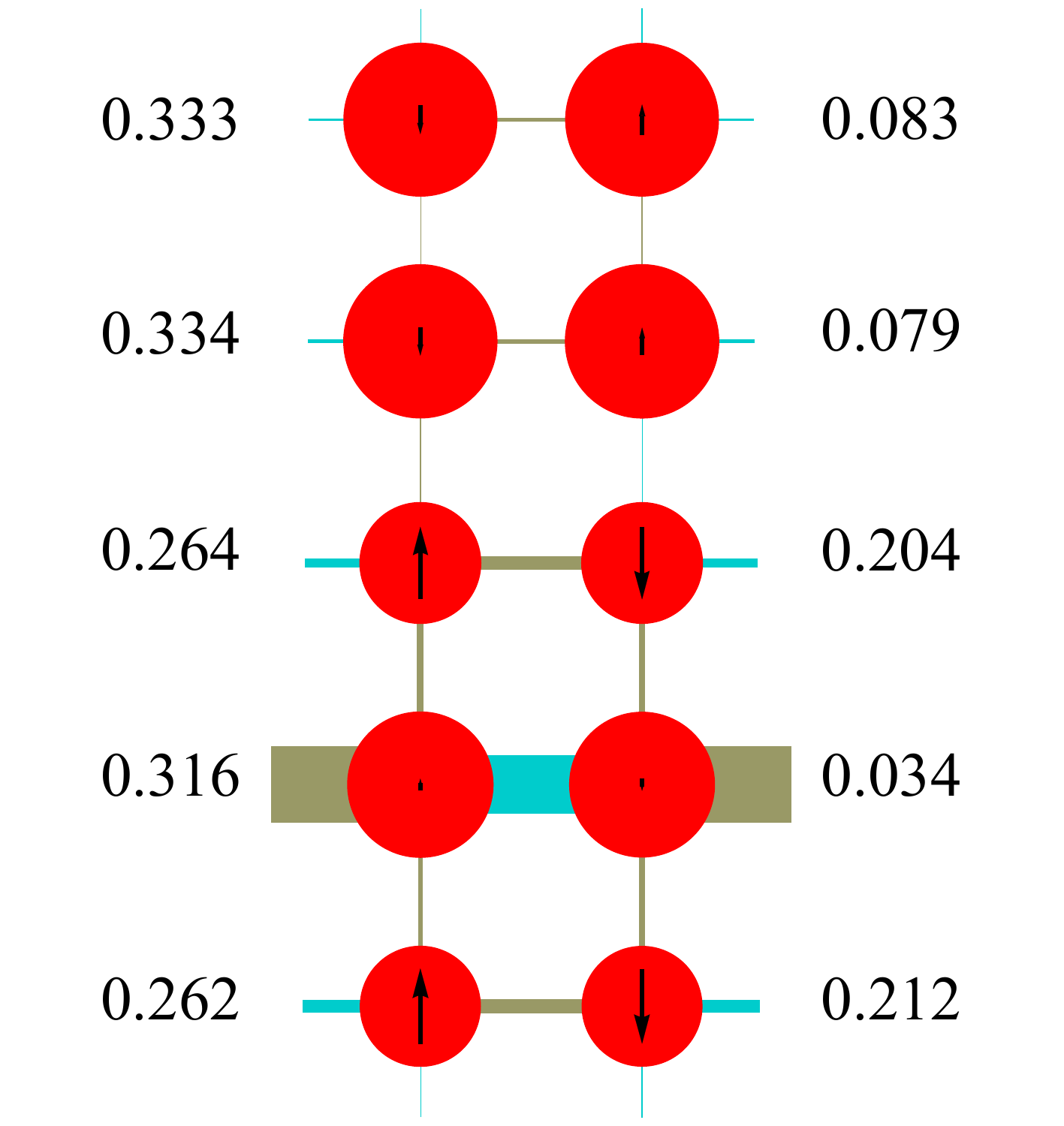}    & 
     \includegraphics[valign=c,width=0.3\textwidth]{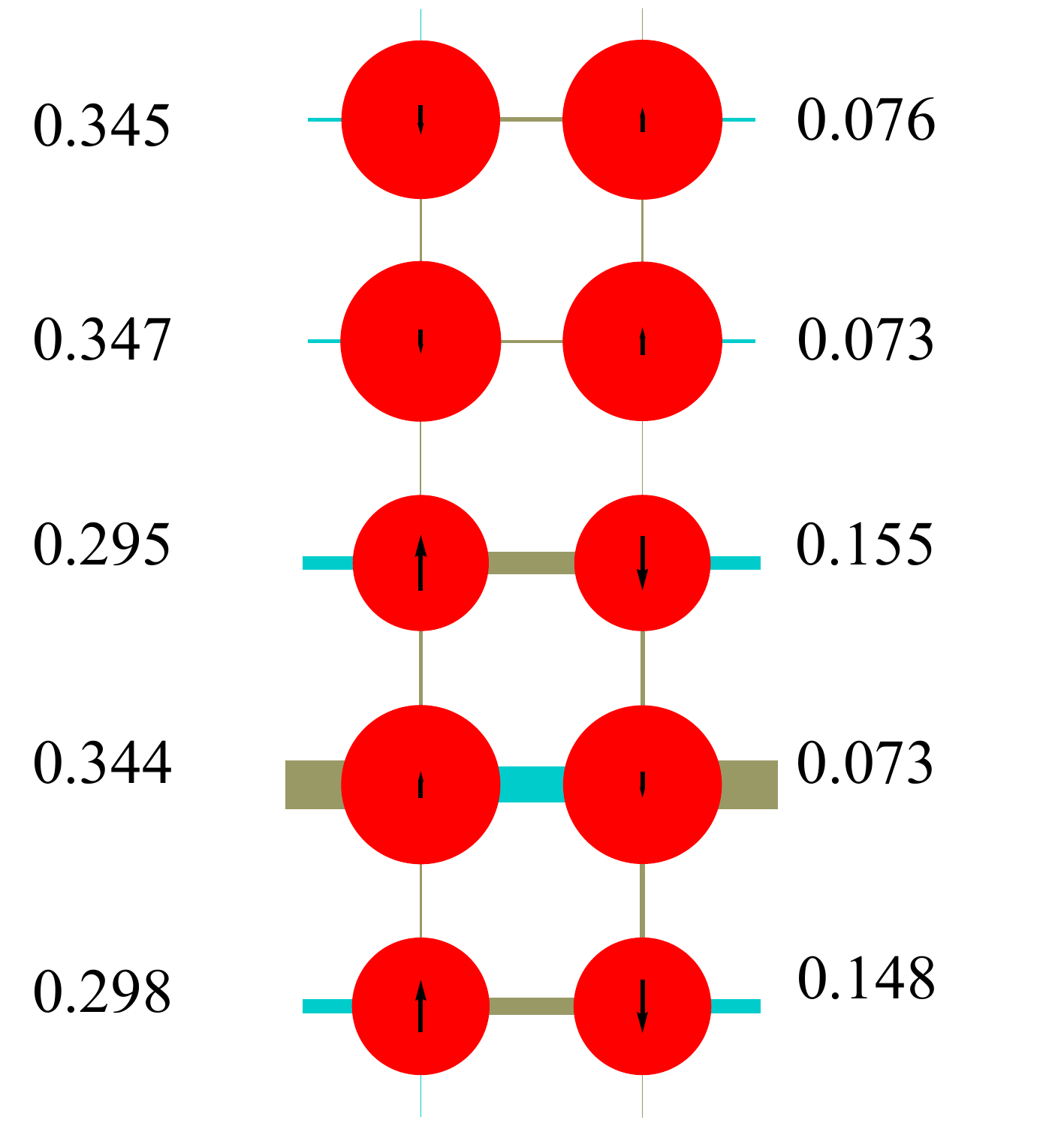} \\
     \hline
     \end{tabular*}
     \label{table:U(1)}
\end{table*}

\begin{table*} [htb!]
  \centering
  \caption{ SU(2) iPEPS with $D^*=6$ and $D^*=8$ at doping $\delta\sim
    0.1$, $0.2$ and $0.3$.  The left column of numbers gives hole
    densities averaged over the two sites in a row.  The number on
    each bond is the singlet pairing amplitude.  } \vspace{3mm}
  \label{tabel:XXX}
     \begin{tabular*}{0.958\textwidth}{|c|c|c|c|}
     \hline
     $ \delta$ & $D^*=6$ & $D^*=6$ & $D^*=8$ \\
     \hline
     $0.1$ & 
     \includegraphics[valign=c,width=0.3\textwidth]{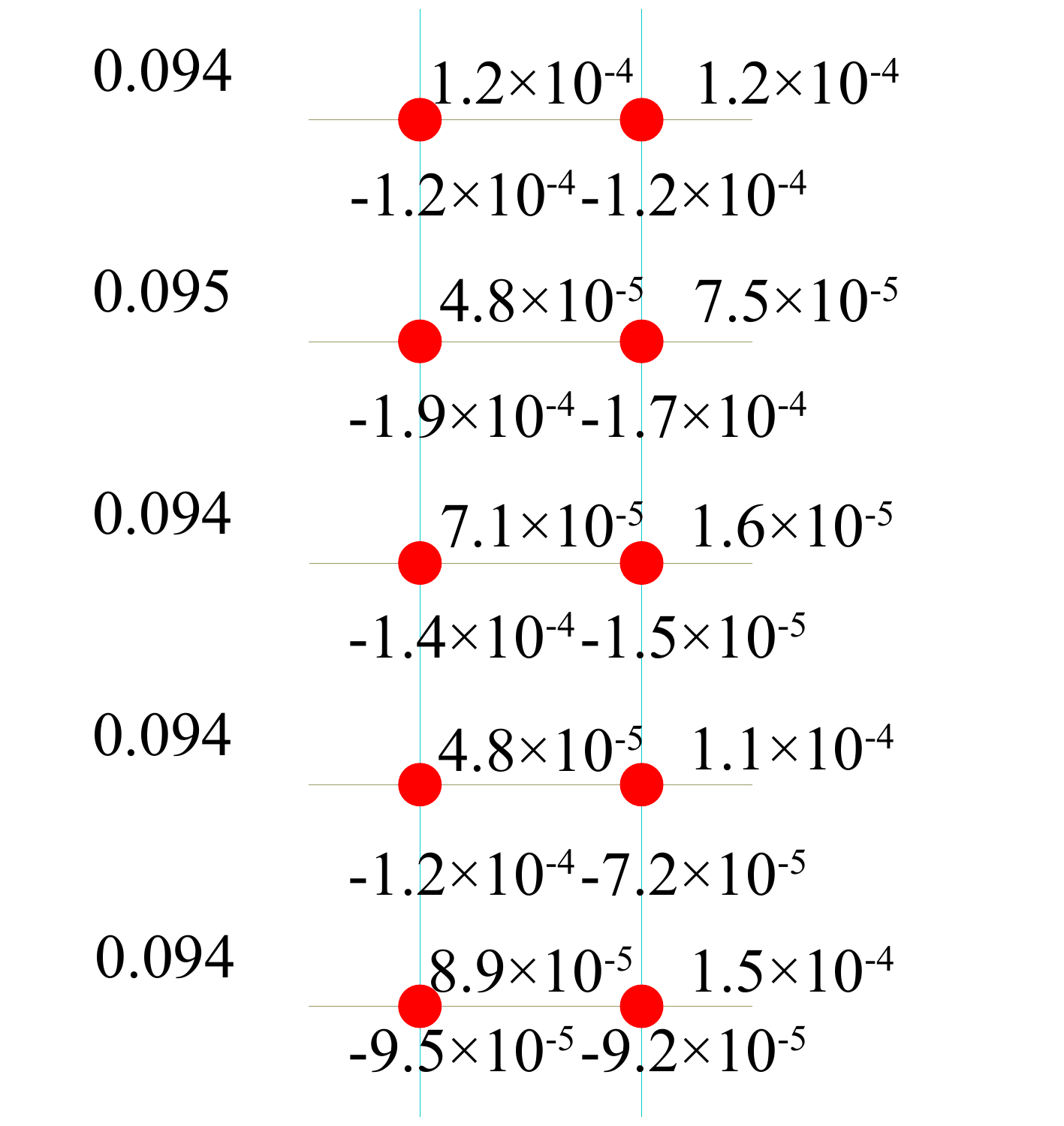} &
     \includegraphics[valign=c,width=0.3\textwidth]{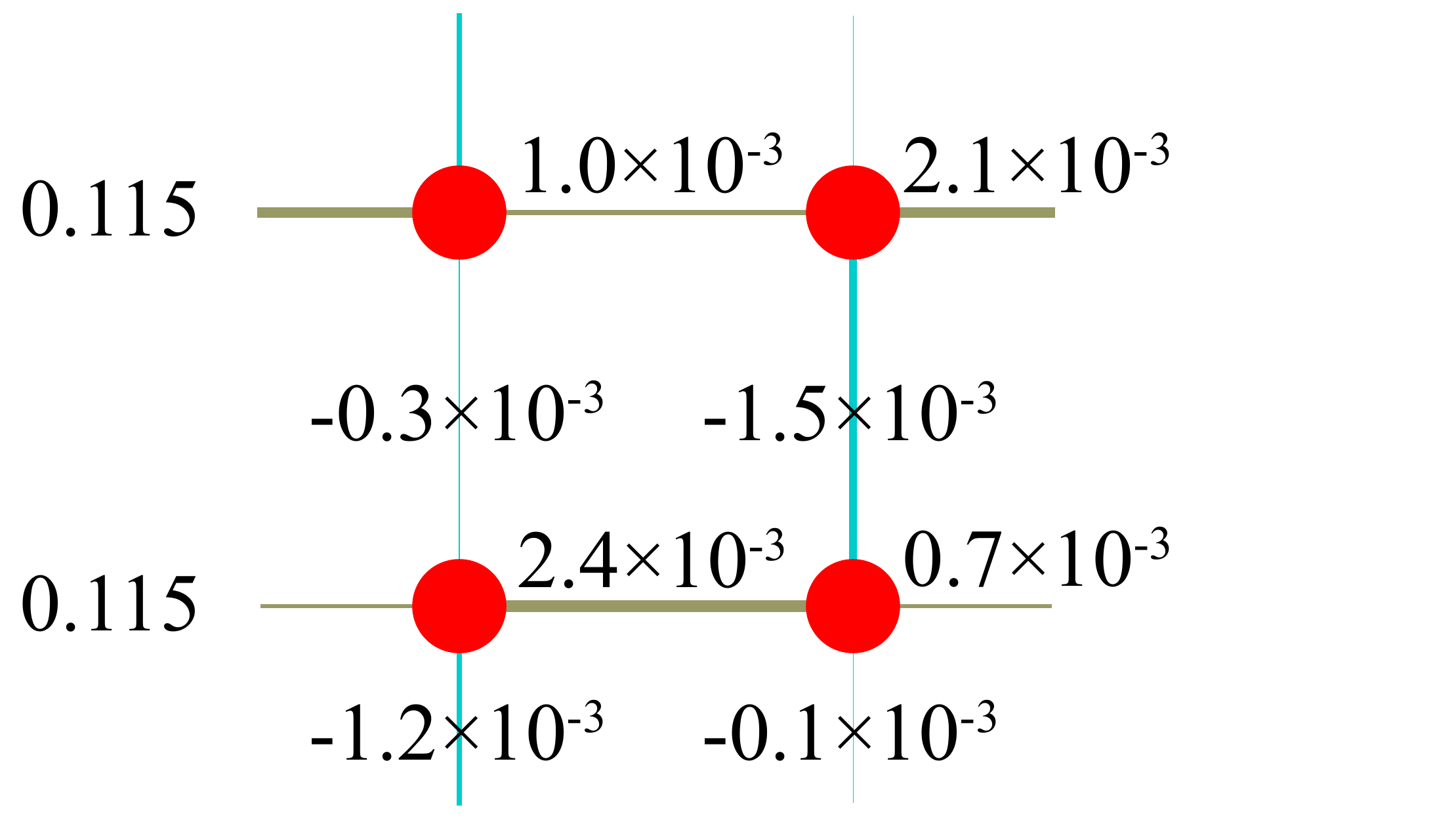} & \\
     \hline
     $0.2$ & 
     \includegraphics[valign=c,width=0.3\textwidth]{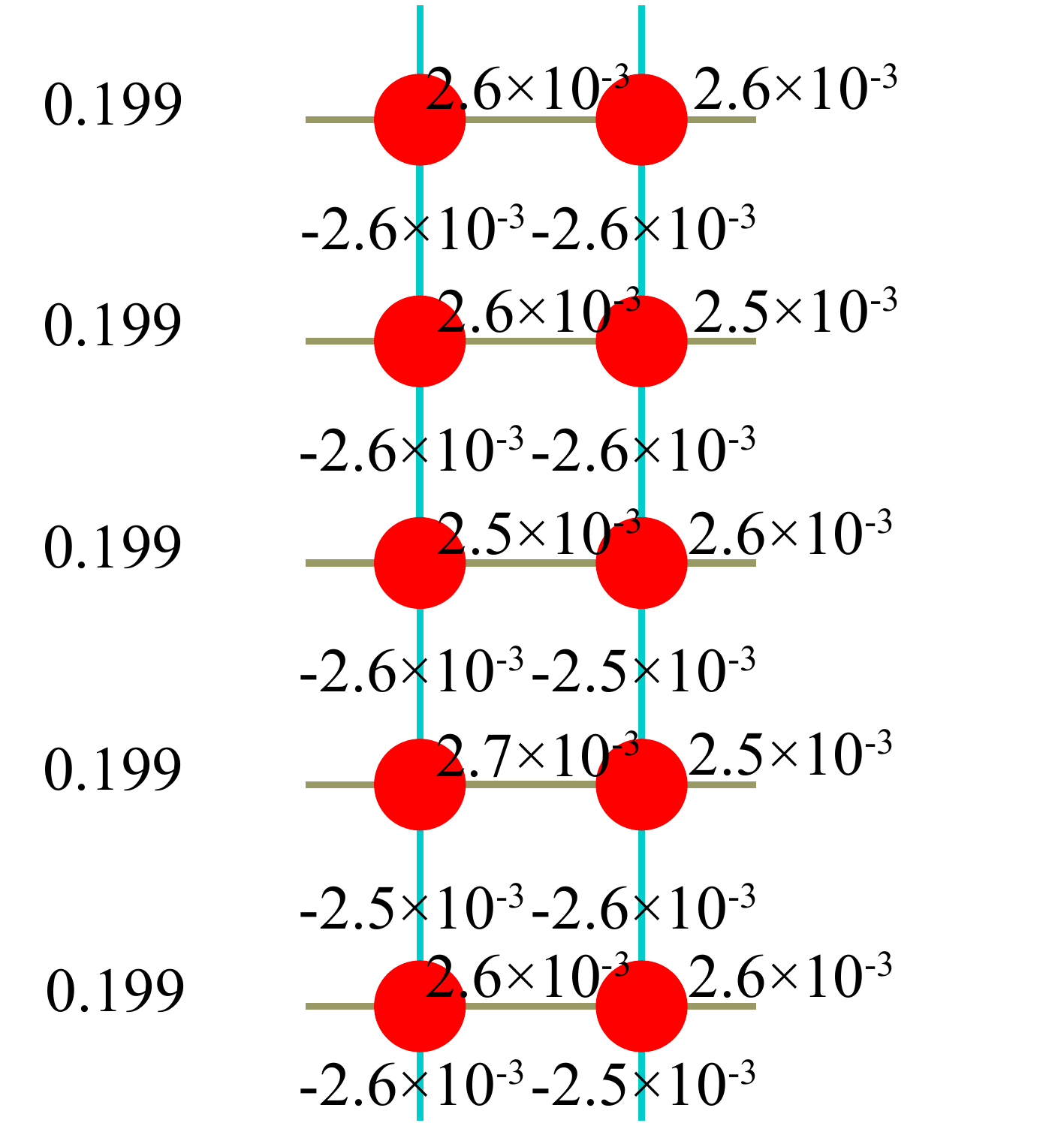} &
     \includegraphics[valign=c,width=0.3\textwidth]{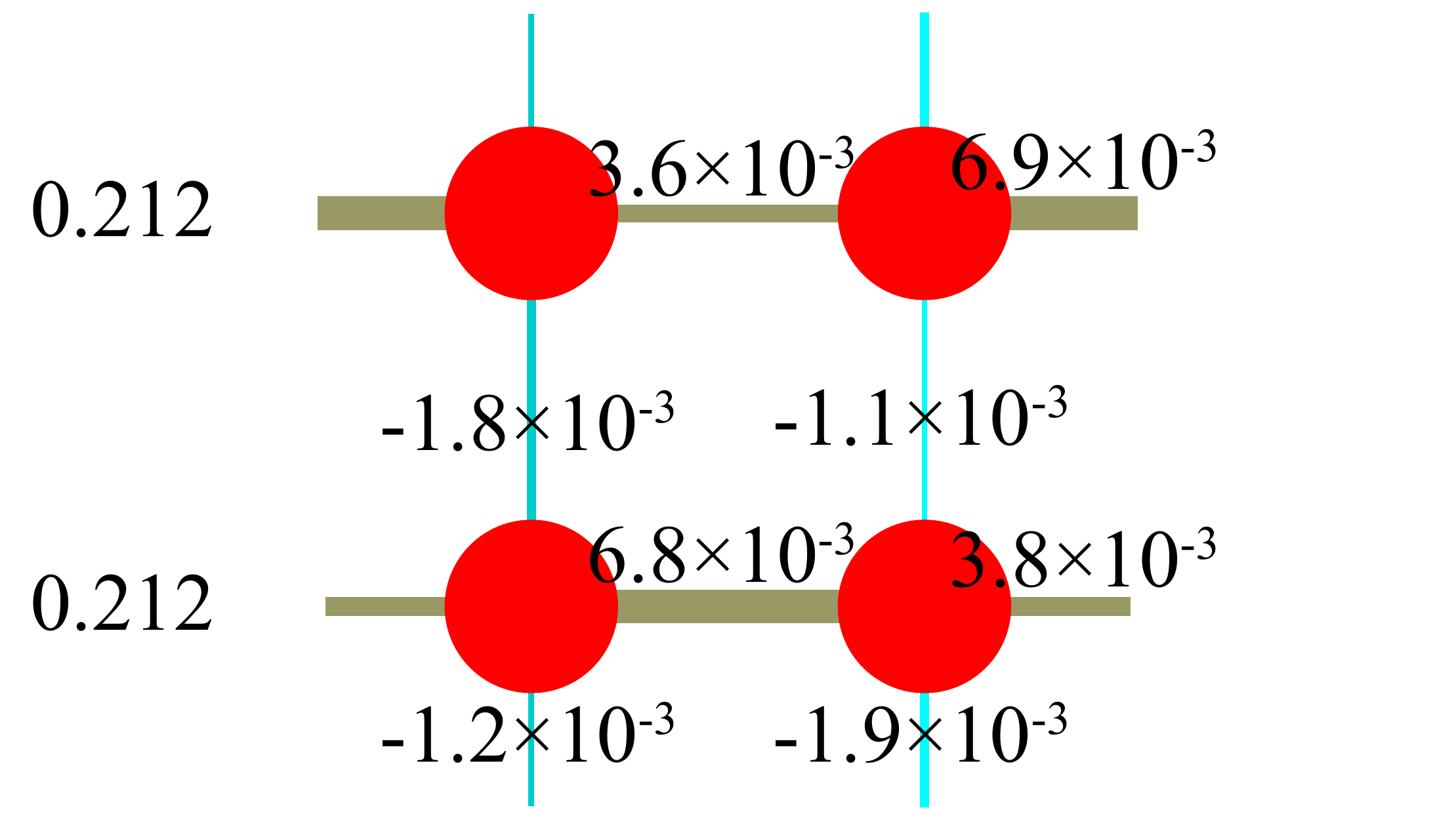} &
     \includegraphics[valign=c,width=0.3\textwidth]{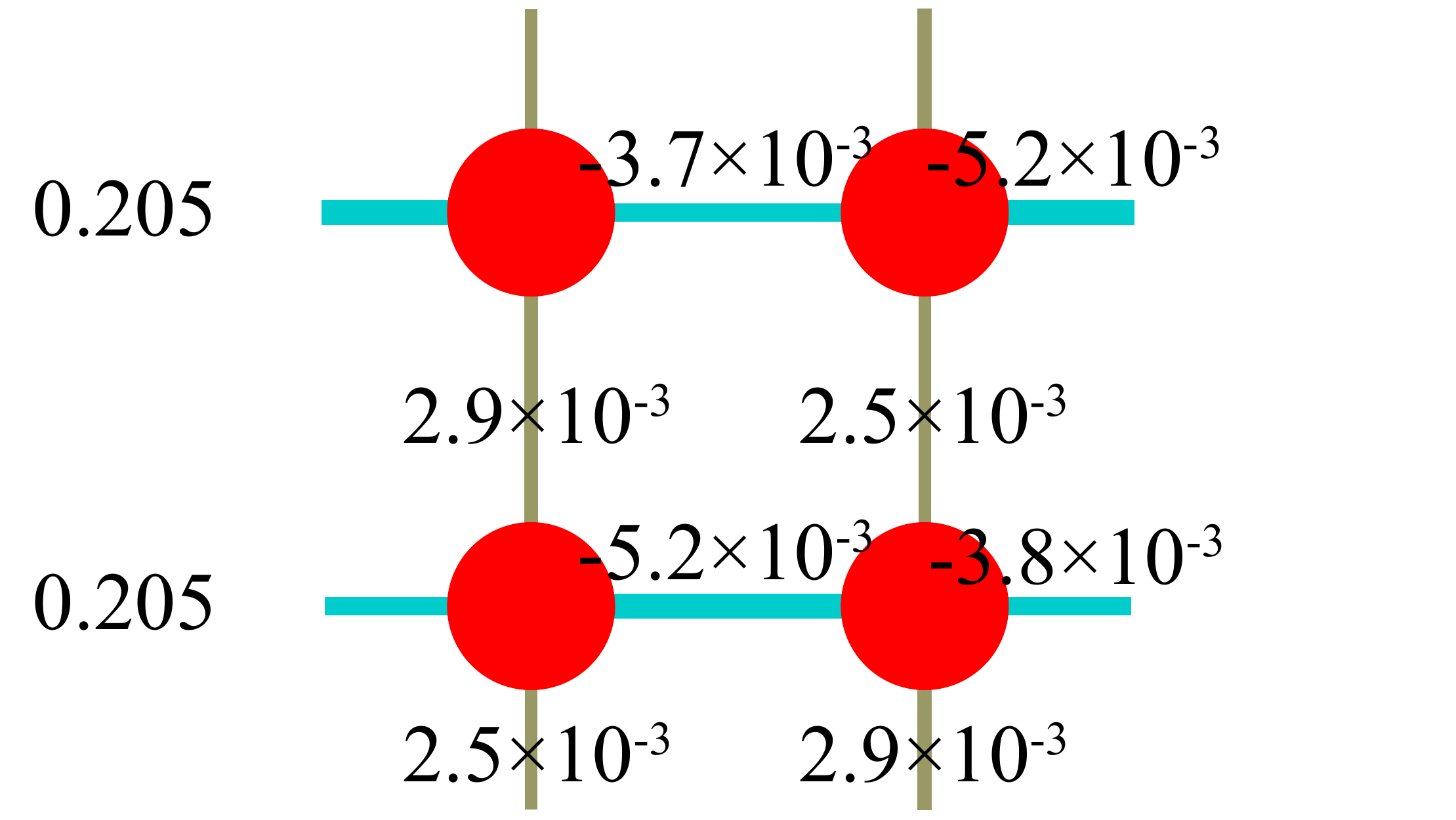} \\
     \hline
     $0.3$ & 
     \includegraphics[valign=c,width=0.3\textwidth]{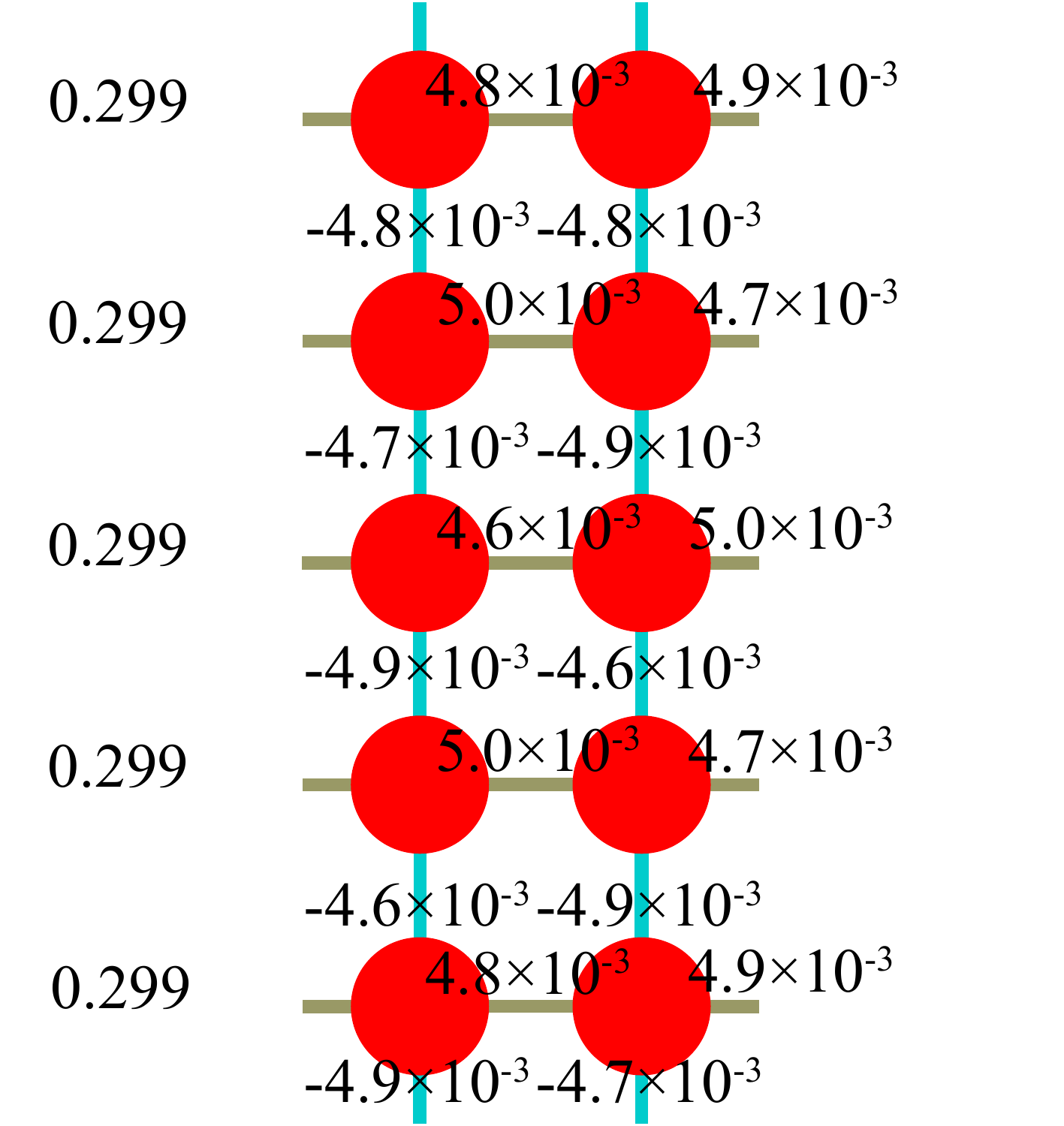} &
     \includegraphics[valign=c,width=0.3\textwidth]{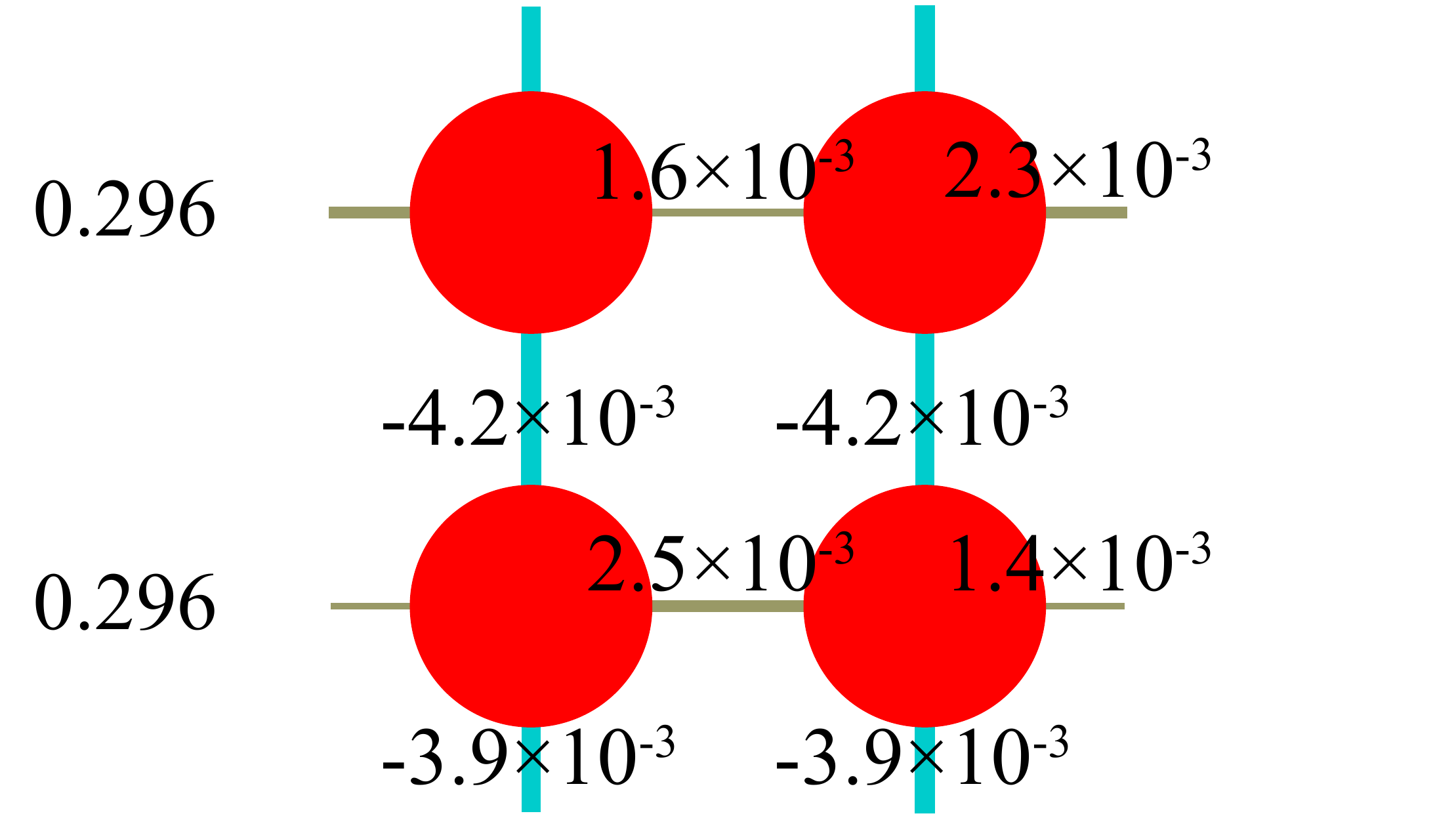} &
     \includegraphics[valign=c,width=0.3\textwidth]{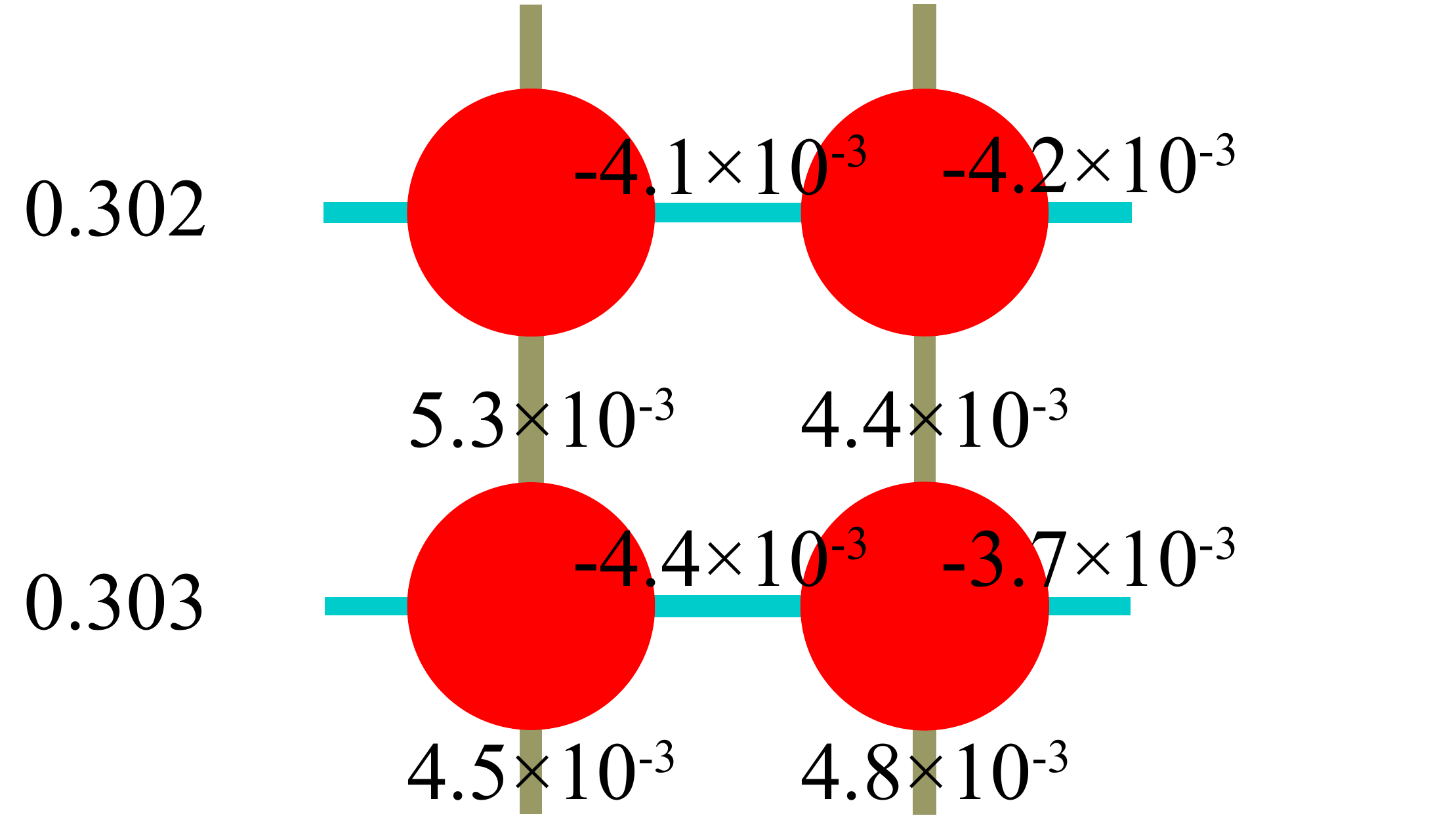} \\
     \hline
     \end{tabular*}
     \label{table:SU(2)}
\end{table*}
\end{document}

%% file: tJ_prb.bbl
%

%% file: tJ_prb.bbl
\begin{thebibliography}{71}%
\makeatletter
\providecommand \@ifxundefined [1]{%
 \@ifx{#1\undefined}
}%
\providecommand \@ifnum [1]{%
 \ifnum #1\expandafter \@firstoftwo
 \else \expandafter \@secondoftwo
 \fi
}%
\providecommand \@ifx [1]{%
 \ifx #1\expandafter \@firstoftwo
 \else \expandafter \@secondoftwo
 \fi
}%
\providecommand \natexlab [1]{#1}%
\providecommand \enquote  [1]{``#1''}%
\providecommand \bibnamefont  [1]{#1}%
\providecommand \bibfnamefont [1]{#1}%
\providecommand \citenamefont [1]{#1}%
\providecommand \href@noop [0]{\@secondoftwo}%
\providecommand \href [0]{\begingroup \@sanitize@url \@href}%
\providecommand \@href[1]{\@@startlink{#1}\@@href}%
\providecommand \@@href[1]{\endgroup#1\@@endlink}%
\providecommand \@sanitize@url [0]{\catcode `\\12\catcode `\$12\catcode
  `\&12\catcode `\#12\catcode `\^12\catcode `\_12\catcode `\%12\relax}%
\providecommand \@@startlink[1]{}%
\providecommand \@@endlink[0]{}%
\providecommand \url  [0]{\begingroup\@sanitize@url \@url }%
\providecommand \@url [1]{\endgroup\@href {#1}{\urlprefix }}%
\providecommand \urlprefix  [0]{URL }%
\providecommand \Eprint [0]{\href }%
\providecommand \doibase [0]{http://dx.doi.org/}%
\providecommand \selectlanguage [0]{\@gobble}%
\providecommand \bibinfo  [0]{\@secondoftwo}%
\providecommand \bibfield  [0]{\@secondoftwo}%
\providecommand \translation [1]{[#1]}%
\providecommand \BibitemOpen [0]{}%
\providecommand \bibitemStop [0]{}%
\providecommand \bibitemNoStop [0]{.\EOS\space}%
\providecommand \EOS [0]{\spacefactor3000\relax}%
\providecommand \BibitemShut  [1]{\csname bibitem#1\endcsname}%
\let\auto@bib@innerbib\@empty
\bibitem [{\citenamefont {Zhang}\ and\ \citenamefont
  {Rice}(1988)}]{Zhang_PRB_1988}%
  \BibitemOpen
  \bibfield  {author} {\bibinfo {author} {\bibfnamefont {F.~C.}\ \bibnamefont
  {Zhang}}\ and\ \bibinfo {author} {\bibfnamefont {T.~M.}\ \bibnamefont
  {Rice}},\ }\href {\doibase 10.1103/PhysRevB.37.3759} {\bibfield  {journal}
  {\bibinfo  {journal} {Phys. Rev. B}\ }\textbf {\bibinfo {volume} {37}},\
  \bibinfo {pages} {3759} (\bibinfo {year} {1988})}\BibitemShut {NoStop}%
\bibitem [{\citenamefont {Poilblanc}\ and\ \citenamefont
  {Rice}(1989)}]{Poilblanc_PRB_1989}%
  \BibitemOpen
  \bibfield  {author} {\bibinfo {author} {\bibfnamefont {D.}~\bibnamefont
  {Poilblanc}}\ and\ \bibinfo {author} {\bibfnamefont {T.~M.}\ \bibnamefont
  {Rice}},\ }\href {\doibase 10.1103/PhysRevB.39.9749} {\bibfield  {journal}
  {\bibinfo  {journal} {Phys. Rev. B}\ }\textbf {\bibinfo {volume} {39}},\
  \bibinfo {pages} {9749} (\bibinfo {year} {1989})}\BibitemShut {NoStop}%
\bibitem [{\citenamefont {Zaanen}\ and\ \citenamefont
  {Gunnarsson}(1989)}]{Zaanen_PRB_1989}%
  \BibitemOpen
  \bibfield  {author} {\bibinfo {author} {\bibfnamefont {J.}~\bibnamefont
  {Zaanen}}\ and\ \bibinfo {author} {\bibfnamefont {O.}~\bibnamefont
  {Gunnarsson}},\ }\href {\doibase 10.1103/PhysRevB.40.7391} {\bibfield
  {journal} {\bibinfo  {journal} {Phys. Rev. B}\ }\textbf {\bibinfo {volume}
  {40}},\ \bibinfo {pages} {7391} (\bibinfo {year} {1989})}\BibitemShut
  {NoStop}%
\bibitem [{\citenamefont {Machida}(1989)}]{Machida_1989}%
  \BibitemOpen
  \bibfield  {author} {\bibinfo {author} {\bibfnamefont {K.}~\bibnamefont
  {Machida}},\ }\href {\doibase 10.1016/0921-4534(89)90316-X} {\bibfield
  {journal} {\bibinfo  {journal} {Physica C: Superconductivity}\ }\textbf
  {\bibinfo {volume} {158}},\ \bibinfo {pages} {192 } (\bibinfo {year}
  {1989})}\BibitemShut {NoStop}%
\bibitem [{\citenamefont {Schulz}(1989)}]{schulz1989domain}%
  \BibitemOpen
  \bibfield  {author} {\bibinfo {author} {\bibfnamefont {H.}~\bibnamefont
  {Schulz}},\ }\href {\doibase 10.1051/jphys:0198900500180283300} {\bibfield
  {journal} {\bibinfo  {journal} {Journal de Physique}\ }\textbf {\bibinfo
  {volume} {50}},\ \bibinfo {pages} {2833} (\bibinfo {year}
  {1989})}\BibitemShut {NoStop}%
\bibitem [{\citenamefont {Emery}\ \emph {et~al.}(1990)\citenamefont {Emery},
  \citenamefont {Kivelson},\ and\ \citenamefont {Lin}}]{Emery_PRL_1990}%
  \BibitemOpen
  \bibfield  {author} {\bibinfo {author} {\bibfnamefont {V.~J.}\ \bibnamefont
  {Emery}}, \bibinfo {author} {\bibfnamefont {S.~A.}\ \bibnamefont {Kivelson}},
  \ and\ \bibinfo {author} {\bibfnamefont {H.~Q.}\ \bibnamefont {Lin}},\ }\href
  {\doibase 10.1103/PhysRevLett.64.475} {\bibfield  {journal} {\bibinfo
  {journal} {Phys. Rev. Lett.}\ }\textbf {\bibinfo {volume} {64}},\ \bibinfo
  {pages} {475} (\bibinfo {year} {1990})}\BibitemShut {NoStop}%
\bibitem [{\citenamefont {Emery}\ and\ \citenamefont
  {Kivelson}(1993)}]{Emery_PC_1993}%
  \BibitemOpen
  \bibfield  {author} {\bibinfo {author} {\bibfnamefont {V.}~\bibnamefont
  {Emery}}\ and\ \bibinfo {author} {\bibfnamefont {S.}~\bibnamefont
  {Kivelson}},\ }\href {https://doi.org/10.1016/0921-4534(93)90581-A}
  {\bibfield  {journal} {\bibinfo  {journal} {Physica C}\ }\textbf {\bibinfo
  {volume} {209}},\ \bibinfo {pages} {597} (\bibinfo {year}
  {1993})}\BibitemShut {NoStop}%
\bibitem [{\citenamefont {Nayak}\ and\ \citenamefont
  {Wilczek}(1997)}]{Chetan_PRL_1997}%
  \BibitemOpen
  \bibfield  {author} {\bibinfo {author} {\bibfnamefont {C.}~\bibnamefont
  {Nayak}}\ and\ \bibinfo {author} {\bibfnamefont {F.}~\bibnamefont
  {Wilczek}},\ }\href {https://doi.org/10.1103/PhysRevLett.78.2465} {\bibfield
  {journal} {\bibinfo  {journal} {Phys. Rev. Lett.}\ }\textbf {\bibinfo
  {volume} {78}},\ \bibinfo {pages} {2465} (\bibinfo {year}
  {1997})}\BibitemShut {NoStop}%
\bibitem [{\citenamefont {White}\ and\ \citenamefont
  {Scalapino}(1998{\natexlab{a}})}]{White_PRL_1998}%
  \BibitemOpen
  \bibfield  {author} {\bibinfo {author} {\bibfnamefont {S.~R.}\ \bibnamefont
  {White}}\ and\ \bibinfo {author} {\bibfnamefont {D.~J.}\ \bibnamefont
  {Scalapino}},\ }\href {\doibase 10.1103/PhysRevLett.80.1272} {\bibfield
  {journal} {\bibinfo  {journal} {Phys. Rev. Lett.}\ }\textbf {\bibinfo
  {volume} {80}},\ \bibinfo {pages} {1272} (\bibinfo {year}
  {1998}{\natexlab{a}})}\BibitemShut {NoStop}%
\bibitem [{\citenamefont {White}\ and\ \citenamefont
  {Scalapino}(1998{\natexlab{b}})}]{White_PRL1998b}%
  \BibitemOpen
  \bibfield  {author} {\bibinfo {author} {\bibfnamefont {S.~R.}\ \bibnamefont
  {White}}\ and\ \bibinfo {author} {\bibfnamefont {D.~J.}\ \bibnamefont
  {Scalapino}},\ }\href {\doibase 10.1103/PhysRevLett.81.3227} {\bibfield
  {journal} {\bibinfo  {journal} {Phys. Rev. Lett.}\ }\textbf {\bibinfo
  {volume} {81}},\ \bibinfo {pages} {3227} (\bibinfo {year}
  {1998}{\natexlab{b}})}\BibitemShut {NoStop}%
\bibitem [{\citenamefont {White}\ and\ \citenamefont
  {Scalapino}(1999)}]{White_PRB_1999}%
  \BibitemOpen
  \bibfield  {author} {\bibinfo {author} {\bibfnamefont {S.~R.}\ \bibnamefont
  {White}}\ and\ \bibinfo {author} {\bibfnamefont {D.~J.}\ \bibnamefont
  {Scalapino}},\ }\href {\doibase 10.1103/PhysRevB.60.R753} {\bibfield
  {journal} {\bibinfo  {journal} {Phys. Rev. B}\ }\textbf {\bibinfo {volume}
  {60}},\ \bibinfo {pages} {R753} (\bibinfo {year} {1999})}\BibitemShut
  {NoStop}%
\bibitem [{\citenamefont {White}\ and\ \citenamefont
  {Scalapino}(2000{\natexlab{a}})}]{White_2000PRB}%
  \BibitemOpen
  \bibfield  {author} {\bibinfo {author} {\bibfnamefont {S.~R.}\ \bibnamefont
  {White}}\ and\ \bibinfo {author} {\bibfnamefont {D.~J.}\ \bibnamefont
  {Scalapino}},\ }\href {\doibase 10.1103/physrevb.61.6320} {\bibfield
  {journal} {\bibinfo  {journal} {Physical Review B}\ }\textbf {\bibinfo
  {volume} {61}},\ \bibinfo {pages} {6320} (\bibinfo {year}
  {2000}{\natexlab{a}})}\BibitemShut {NoStop}%
\bibitem [{\citenamefont {Eskes}\ \emph {et~al.}(1998)\citenamefont {Eskes},
  \citenamefont {Osman}, \citenamefont {Grimberg}, \citenamefont {van
  Saarloos},\ and\ \citenamefont {Zaanen}}]{Eskes_PRB_1998}%
  \BibitemOpen
  \bibfield  {author} {\bibinfo {author} {\bibfnamefont {H.}~\bibnamefont
  {Eskes}}, \bibinfo {author} {\bibfnamefont {O.~Y.}\ \bibnamefont {Osman}},
  \bibinfo {author} {\bibfnamefont {R.}~\bibnamefont {Grimberg}}, \bibinfo
  {author} {\bibfnamefont {W.}~\bibnamefont {van Saarloos}}, \ and\ \bibinfo
  {author} {\bibfnamefont {J.}~\bibnamefont {Zaanen}},\ }\href
  {https://doi.org/10.1103/PhysRevB.58.6963} {\bibfield  {journal} {\bibinfo
  {journal} {Phys. Rev. B}\ }\textbf {\bibinfo {volume} {58}},\ \bibinfo
  {pages} {6963} (\bibinfo {year} {1998})}\BibitemShut {NoStop}%
\bibitem [{\citenamefont {Pryadko}\ \emph {et~al.}(1999)\citenamefont
  {Pryadko}, \citenamefont {Kivelson}, \citenamefont {Emery}, \citenamefont
  {Bazaliy},\ and\ \citenamefont {Demler}}]{Pryadko_PRB_1999}%
  \BibitemOpen
  \bibfield  {author} {\bibinfo {author} {\bibfnamefont {L.~P.}\ \bibnamefont
  {Pryadko}}, \bibinfo {author} {\bibfnamefont {S.~A.}\ \bibnamefont
  {Kivelson}}, \bibinfo {author} {\bibfnamefont {V.~J.}\ \bibnamefont {Emery}},
  \bibinfo {author} {\bibfnamefont {Y.~B.}\ \bibnamefont {Bazaliy}}, \ and\
  \bibinfo {author} {\bibfnamefont {E.~A.}\ \bibnamefont {Demler}},\ }\href
  {https://doi.org/10.1103/PhysRevB.60.7541} {\bibfield  {journal} {\bibinfo
  {journal} {Phys. Rev. B}\ }\textbf {\bibinfo {volume} {60}},\ \bibinfo
  {pages} {7541} (\bibinfo {year} {1999})}\BibitemShut {NoStop}%
\bibitem [{\citenamefont {White}\ and\ \citenamefont
  {Scalapino}(2000{\natexlab{b}})}]{White_PRB2000}%
  \BibitemOpen
  \bibfield  {author} {\bibinfo {author} {\bibfnamefont {S.~R.}\ \bibnamefont
  {White}}\ and\ \bibinfo {author} {\bibfnamefont {D.~J.}\ \bibnamefont
  {Scalapino}},\ }\href {\doibase 10.1103/PhysRevB.61.6320} {\bibfield
  {journal} {\bibinfo  {journal} {Phys. Rev. B}\ }\textbf {\bibinfo {volume}
  {61}},\ \bibinfo {pages} {6320} (\bibinfo {year}
  {2000}{\natexlab{b}})}\BibitemShut {NoStop}%
\bibitem [{\citenamefont {Chernyshev}\ \emph {et~al.}(2002)\citenamefont
  {Chernyshev}, \citenamefont {White},\ and\ \citenamefont
  {Castro~Neto}}]{White_PRB2002}%
  \BibitemOpen
  \bibfield  {author} {\bibinfo {author} {\bibfnamefont {A.~L.}\ \bibnamefont
  {Chernyshev}}, \bibinfo {author} {\bibfnamefont {S.~R.}\ \bibnamefont
  {White}}, \ and\ \bibinfo {author} {\bibfnamefont {A.~H.}\ \bibnamefont
  {Castro~Neto}},\ }\href {\doibase 10.1103/PhysRevB.65.214527} {\bibfield
  {journal} {\bibinfo  {journal} {Phys. Rev. B}\ }\textbf {\bibinfo {volume}
  {65}},\ \bibinfo {pages} {214527} (\bibinfo {year} {2002})}\BibitemShut
  {NoStop}%
\bibitem [{\citenamefont {Himeda}\ \emph {et~al.}(2002)\citenamefont {Himeda},
  \citenamefont {Kato},\ and\ \citenamefont {Ogata}}]{Himeda_2002_PRL}%
  \BibitemOpen
  \bibfield  {author} {\bibinfo {author} {\bibfnamefont {A.}~\bibnamefont
  {Himeda}}, \bibinfo {author} {\bibfnamefont {T.}~\bibnamefont {Kato}}, \ and\
  \bibinfo {author} {\bibfnamefont {M.}~\bibnamefont {Ogata}},\ }\href
  {\doibase 10.1103/PhysRevLett.88.117001} {\bibfield  {journal} {\bibinfo
  {journal} {Phys. Rev. Lett.}\ }\textbf {\bibinfo {volume} {88}},\ \bibinfo
  {pages} {117001} (\bibinfo {year} {2002})}\BibitemShut {NoStop}%
\bibitem [{\citenamefont {Chou}\ \emph
  {et~al.}(2008{\natexlab{a}})\citenamefont {Chou}, \citenamefont {Fukushima},\
  and\ \citenamefont {Lee}}]{CCP_PRB_2008}%
  \BibitemOpen
  \bibfield  {author} {\bibinfo {author} {\bibfnamefont {C.-P.}\ \bibnamefont
  {Chou}}, \bibinfo {author} {\bibfnamefont {N.}~\bibnamefont {Fukushima}}, \
  and\ \bibinfo {author} {\bibfnamefont {T.~K.}\ \bibnamefont {Lee}},\ }\href
  {\doibase 10.1103/PhysRevB.78.134530} {\bibfield  {journal} {\bibinfo
  {journal} {Phys. Rev. B}\ }\textbf {\bibinfo {volume} {78}},\ \bibinfo
  {pages} {134530} (\bibinfo {year} {2008}{\natexlab{a}})}\BibitemShut
  {NoStop}%
\bibitem [{\citenamefont {Yang}\ \emph {et~al.}(2009)\citenamefont {Yang},
  \citenamefont {Chen}, \citenamefont {Rice}, \citenamefont {Sigrist},\ and\
  \citenamefont {Zhang}}]{Yang_2009}%
  \BibitemOpen
  \bibfield  {author} {\bibinfo {author} {\bibfnamefont {K.-Y.}\ \bibnamefont
  {Yang}}, \bibinfo {author} {\bibfnamefont {W.~Q.}\ \bibnamefont {Chen}},
  \bibinfo {author} {\bibfnamefont {T.~M.}\ \bibnamefont {Rice}}, \bibinfo
  {author} {\bibfnamefont {M.}~\bibnamefont {Sigrist}}, \ and\ \bibinfo
  {author} {\bibfnamefont {F.-C.}\ \bibnamefont {Zhang}},\ }\href {\doibase
  10.1088/1367-2630/11/5/055053} {\bibfield  {journal} {\bibinfo  {journal}
  {New Journal of Physics}\ }\textbf {\bibinfo {volume} {11}},\ \bibinfo
  {pages} {055053} (\bibinfo {year} {2009})}\BibitemShut {NoStop}%
\bibitem [{\citenamefont {Corboz}\ \emph {et~al.}(2011)\citenamefont {Corboz},
  \citenamefont {White}, \citenamefont {Vidal},\ and\ \citenamefont
  {Troyer}}]{Corboz_PRBR_2011}%
  \BibitemOpen
  \bibfield  {author} {\bibinfo {author} {\bibfnamefont {P.}~\bibnamefont
  {Corboz}}, \bibinfo {author} {\bibfnamefont {S.~R.}\ \bibnamefont {White}},
  \bibinfo {author} {\bibfnamefont {G.}~\bibnamefont {Vidal}}, \ and\ \bibinfo
  {author} {\bibfnamefont {M.}~\bibnamefont {Troyer}},\ }\href {\doibase
  10.1103/PhysRevB.84.041108} {\bibfield  {journal} {\bibinfo  {journal} {Phys.
  Rev. B}\ }\textbf {\bibinfo {volume} {84}},\ \bibinfo {pages} {041108}
  (\bibinfo {year} {2011})}\BibitemShut {NoStop}%
\bibitem [{\citenamefont {Sorella}\ \emph {et~al.}(2002)\citenamefont
  {Sorella}, \citenamefont {Martins}, \citenamefont {Becca}, \citenamefont
  {Gazza}, \citenamefont {Capriotti}, \citenamefont {Parola},\ and\
  \citenamefont {Dagotto}}]{Sorella_PRL_2012}%
  \BibitemOpen
  \bibfield  {author} {\bibinfo {author} {\bibfnamefont {S.}~\bibnamefont
  {Sorella}}, \bibinfo {author} {\bibfnamefont {G.~B.}\ \bibnamefont
  {Martins}}, \bibinfo {author} {\bibfnamefont {F.}~\bibnamefont {Becca}},
  \bibinfo {author} {\bibfnamefont {C.}~\bibnamefont {Gazza}}, \bibinfo
  {author} {\bibfnamefont {L.}~\bibnamefont {Capriotti}}, \bibinfo {author}
  {\bibfnamefont {A.}~\bibnamefont {Parola}}, \ and\ \bibinfo {author}
  {\bibfnamefont {E.}~\bibnamefont {Dagotto}},\ }\href {\doibase
  10.1103/PhysRevLett.88.117002} {\bibfield  {journal} {\bibinfo  {journal}
  {Phys. Rev. Lett.}\ }\textbf {\bibinfo {volume} {88}},\ \bibinfo {pages}
  {117002} (\bibinfo {year} {2002})}\BibitemShut {NoStop}%
\bibitem [{\citenamefont {Hu}\ \emph {et~al.}(2012)\citenamefont {Hu},
  \citenamefont {Becca},\ and\ \citenamefont {Sorella}}]{HW_PRB_2012}%
  \BibitemOpen
  \bibfield  {author} {\bibinfo {author} {\bibfnamefont {W.-J.}\ \bibnamefont
  {Hu}}, \bibinfo {author} {\bibfnamefont {F.}~\bibnamefont {Becca}}, \ and\
  \bibinfo {author} {\bibfnamefont {S.}~\bibnamefont {Sorella}},\ }\href
  {\doibase 10.1103/PhysRevB.85.081110} {\bibfield  {journal} {\bibinfo
  {journal} {Phys. Rev. B}\ }\textbf {\bibinfo {volume} {85}},\ \bibinfo
  {pages} {081110} (\bibinfo {year} {2012})}\BibitemShut {NoStop}%
\bibitem [{\citenamefont {Corboz}\ \emph {et~al.}(2014)\citenamefont {Corboz},
  \citenamefont {Rice},\ and\ \citenamefont {Troyer}}]{Corboz_PRL_2014}%
  \BibitemOpen
  \bibfield  {author} {\bibinfo {author} {\bibfnamefont {P.}~\bibnamefont
  {Corboz}}, \bibinfo {author} {\bibfnamefont {T.~M.}\ \bibnamefont {Rice}}, \
  and\ \bibinfo {author} {\bibfnamefont {M.}~\bibnamefont {Troyer}},\ }\href
  {\doibase 10.1103/PhysRevLett.113.046402} {\bibfield  {journal} {\bibinfo
  {journal} {Phys. Rev. Lett.}\ }\textbf {\bibinfo {volume} {113}},\ \bibinfo
  {pages} {046402} (\bibinfo {year} {2014})}\BibitemShut {NoStop}%
\bibitem [{\citenamefont {Hellberg}\ and\ \citenamefont
  {Manousakis}(1999)}]{Hellberg_PRL_1999}%
  \BibitemOpen
  \bibfield  {author} {\bibinfo {author} {\bibfnamefont {C.~S.}\ \bibnamefont
  {Hellberg}}\ and\ \bibinfo {author} {\bibfnamefont {E.}~\bibnamefont
  {Manousakis}},\ }\href {https://doi.org/10.1103/PhysRevLett.83.132}
  {\bibfield  {journal} {\bibinfo  {journal} {Phys. Rev. Lett.}\ }\textbf
  {\bibinfo {volume} {83}},\ \bibinfo {pages} {132} (\bibinfo {year}
  {1999})}\BibitemShut {NoStop}%
\bibitem [{\citenamefont {Raczkowski}\ \emph {et~al.}(2007)\citenamefont
  {Raczkowski}, \citenamefont {Capello}, \citenamefont {Poilblanc},
  \citenamefont {Fr\'esard},\ and\ \citenamefont {Ole\ifmmode~\acute{s}\else
  \'{s}\fi{}}}]{Raczkowski_PRB_2007}%
  \BibitemOpen
  \bibfield  {author} {\bibinfo {author} {\bibfnamefont {M.}~\bibnamefont
  {Raczkowski}}, \bibinfo {author} {\bibfnamefont {M.}~\bibnamefont {Capello}},
  \bibinfo {author} {\bibfnamefont {D.}~\bibnamefont {Poilblanc}}, \bibinfo
  {author} {\bibfnamefont {R.}~\bibnamefont {Fr\'esard}}, \ and\ \bibinfo
  {author} {\bibfnamefont {A.~M.}\ \bibnamefont {Ole\ifmmode~\acute{s}\else
  \'{s}\fi{}}},\ }\href {\doibase 10.1103/PhysRevB.76.140505} {\bibfield
  {journal} {\bibinfo  {journal} {Phys. Rev. B}\ }\textbf {\bibinfo {volume}
  {76}},\ \bibinfo {pages} {140505} (\bibinfo {year} {2007})}\BibitemShut
  {NoStop}%
\bibitem [{\citenamefont {Chou}\ \emph
  {et~al.}(2008{\natexlab{b}})\citenamefont {Chou}, \citenamefont {Fukushima},\
  and\ \citenamefont {Lee}}]{Chou_PRB_2008}%
  \BibitemOpen
  \bibfield  {author} {\bibinfo {author} {\bibfnamefont {C.-P.}\ \bibnamefont
  {Chou}}, \bibinfo {author} {\bibfnamefont {N.}~\bibnamefont {Fukushima}}, \
  and\ \bibinfo {author} {\bibfnamefont {T.~K.}\ \bibnamefont {Lee}},\ }\href
  {\doibase 10.1103/PhysRevB.78.134530} {\bibfield  {journal} {\bibinfo
  {journal} {Phys. Rev. B}\ }\textbf {\bibinfo {volume} {78}},\ \bibinfo
  {pages} {134530} (\bibinfo {year} {2008}{\natexlab{b}})}\BibitemShut
  {NoStop}%
\bibitem [{\citenamefont {Zheng}\ \emph {et~al.}(2017)\citenamefont {Zheng},
  \citenamefont {Chung}, \citenamefont {Corboz}, \citenamefont {Ehlers},
  \citenamefont {Qin}, \citenamefont {Noack}, \citenamefont {Shi},
  \citenamefont {White}, \citenamefont {Zhang},\ and\ \citenamefont
  {Chan}}]{Zheng_Science_2017}%
  \BibitemOpen
  \bibfield  {author} {\bibinfo {author} {\bibfnamefont {B.-X.}\ \bibnamefont
  {Zheng}}, \bibinfo {author} {\bibfnamefont {C.-M.}\ \bibnamefont {Chung}},
  \bibinfo {author} {\bibfnamefont {P.}~\bibnamefont {Corboz}}, \bibinfo
  {author} {\bibfnamefont {G.}~\bibnamefont {Ehlers}}, \bibinfo {author}
  {\bibfnamefont {M.-P.}\ \bibnamefont {Qin}}, \bibinfo {author} {\bibfnamefont
  {R.~M.}\ \bibnamefont {Noack}}, \bibinfo {author} {\bibfnamefont
  {H.}~\bibnamefont {Shi}}, \bibinfo {author} {\bibfnamefont {S.~R.}\
  \bibnamefont {White}}, \bibinfo {author} {\bibfnamefont {S.}~\bibnamefont
  {Zhang}}, \ and\ \bibinfo {author} {\bibfnamefont {G.~K.-L.}\ \bibnamefont
  {Chan}},\ }\href {https://doi.org/10.1126/science.aam7127} {\bibfield
  {journal} {\bibinfo  {journal} {Science}\ }\textbf {\bibinfo {volume}
  {358}},\ \bibinfo {pages} {1155} (\bibinfo {year} {2017})}\BibitemShut
  {NoStop}%
\bibitem [{\citenamefont {Huang}\ \emph {et~al.}(2018)\citenamefont {Huang},
  \citenamefont {Mendl}, \citenamefont {Jiang}, \citenamefont {Moritz},\ and\
  \citenamefont {Devereaux}}]{Huang_Nature_2018}%
  \BibitemOpen
  \bibfield  {author} {\bibinfo {author} {\bibfnamefont {E.~W.}\ \bibnamefont
  {Huang}}, \bibinfo {author} {\bibfnamefont {C.~B.}\ \bibnamefont {Mendl}},
  \bibinfo {author} {\bibfnamefont {H.-C.}\ \bibnamefont {Jiang}}, \bibinfo
  {author} {\bibfnamefont {B.}~\bibnamefont {Moritz}}, \ and\ \bibinfo {author}
  {\bibfnamefont {T.~P.}\ \bibnamefont {Devereaux}},\ }\href
  {https://doi.org/10.1038/s41535-018-0097-0} {\bibfield  {journal} {\bibinfo
  {journal} {npj Quantum Materials}\ }\textbf {\bibinfo {volume} {3}},\
  \bibinfo {pages} {22} (\bibinfo {year} {2018})}\BibitemShut {NoStop}%
\bibitem [{\citenamefont {Ido}\ \emph {et~al.}(2018)\citenamefont {Ido},
  \citenamefont {Ohgoe},\ and\ \citenamefont {Imada}}]{Ido_PRB_2018}%
  \BibitemOpen
  \bibfield  {author} {\bibinfo {author} {\bibfnamefont {K.}~\bibnamefont
  {Ido}}, \bibinfo {author} {\bibfnamefont {T.}~\bibnamefont {Ohgoe}}, \ and\
  \bibinfo {author} {\bibfnamefont {M.}~\bibnamefont {Imada}},\ }\href
  {https://doi.org/10.1103/PhysRevB.97.045138} {\bibfield  {journal} {\bibinfo
  {journal} {Phys. Rev. B}\ }\textbf {\bibinfo {volume} {97}},\ \bibinfo
  {pages} {045138} (\bibinfo {year} {2018})}\BibitemShut {NoStop}%
\bibitem [{\citenamefont {Darmawan}\ \emph {et~al.}(2018)\citenamefont
  {Darmawan}, \citenamefont {Nomura}, \citenamefont {Yamaji},\ and\
  \citenamefont {Imada}}]{Darmawan_PRB_2018}%
  \BibitemOpen
  \bibfield  {author} {\bibinfo {author} {\bibfnamefont {A.~S.}\ \bibnamefont
  {Darmawan}}, \bibinfo {author} {\bibfnamefont {Y.}~\bibnamefont {Nomura}},
  \bibinfo {author} {\bibfnamefont {Y.}~\bibnamefont {Yamaji}}, \ and\ \bibinfo
  {author} {\bibfnamefont {M.}~\bibnamefont {Imada}},\ }\href
  {https://doi.org/10.1103/PhysRevB.98.205132} {\bibfield  {journal} {\bibinfo
  {journal} {Phys. Rev. B}\ }\textbf {\bibinfo {volume} {98}},\ \bibinfo
  {pages} {205132} (\bibinfo {year} {2018})}\BibitemShut {NoStop}%
\bibitem [{\citenamefont {Jiang}\ and\ \citenamefont
  {Devereaux}(2019)}]{Jiang2019}%
  \BibitemOpen
  \bibfield  {author} {\bibinfo {author} {\bibfnamefont {H.-C.}\ \bibnamefont
  {Jiang}}\ and\ \bibinfo {author} {\bibfnamefont {T.~P.}\ \bibnamefont
  {Devereaux}},\ }\href {\doibase 10.1126/science.aal5304} {\bibfield
  {journal} {\bibinfo  {journal} {Science}\ }\textbf {\bibinfo {volume}
  {365}},\ \bibinfo {pages} {1424} (\bibinfo {year} {2019})}\BibitemShut
  {NoStop}%
\bibitem [{\citenamefont {Ponsioen}\ \emph {et~al.}()\citenamefont {Ponsioen},
  \citenamefont {Chung},\ and\ \citenamefont {Corboz}}]{Ponsioen2019}%
  \BibitemOpen
  \bibfield  {author} {\bibinfo {author} {\bibfnamefont {B.}~\bibnamefont
  {Ponsioen}}, \bibinfo {author} {\bibfnamefont {S.~S.}\ \bibnamefont {Chung}},
  \ and\ \bibinfo {author} {\bibfnamefont {P.}~\bibnamefont {Corboz}},\ }\href
  {http://arxiv.org/abs/1907.01909v1} {\bibinfo  {journal} {arXiv:1907.01909}\
  }\BibitemShut {NoStop}%
\bibitem [{\citenamefont {Mingpu~Qin}\ and\ \citenamefont
  {Zhang}(2019)}]{Qin2019}%
  \BibitemOpen
\bibfield  {journal} {  }\bibfield  {author} {\bibinfo {author} {\bibfnamefont
  {H.~S. E. V. C. H. U. S. S. R.~W.}\ \bibnamefont {Mingpu~Qin}, \bibfnamefont
  {Chia-Min~Chung}}\ and\ \bibinfo {author} {\bibfnamefont {S.}~\bibnamefont
  {Zhang}},\ }\href {https://arxiv.org/abs/1910.08931} {\bibfield  {journal}
  {\bibinfo  {journal} {arXiv:1910.08931 [cond-mat.str-el]}\ } (\bibinfo {year}
  {2019})}\BibitemShut {NoStop}%
\bibitem [{\citenamefont {Jiang}\ \emph {et~al.}(2018)\citenamefont {Jiang},
  \citenamefont {Weng},\ and\ \citenamefont {Kivelson}}]{Jiang_PRB_2018}%
  \BibitemOpen
  \bibfield  {author} {\bibinfo {author} {\bibfnamefont {H.-C.}\ \bibnamefont
  {Jiang}}, \bibinfo {author} {\bibfnamefont {Z.-Y.}\ \bibnamefont {Weng}}, \
  and\ \bibinfo {author} {\bibfnamefont {S.~A.}\ \bibnamefont {Kivelson}},\
  }\href {https://doi.org/10.1103/PhysRevB.98.140505} {\bibfield  {journal}
  {\bibinfo  {journal} {Phys. Rev. B}\ }\textbf {\bibinfo {volume} {98}},\
  \bibinfo {pages} {140505} (\bibinfo {year} {2018})}\BibitemShut {NoStop}%
\bibitem [{\citenamefont {Wang}\ \emph {et~al.}(2013)\citenamefont {Wang},
  \citenamefont {Poilblanc}, \citenamefont {Gu}, \citenamefont {Wen},\ and\
  \citenamefont {Verstraete}}]{Wang_PRL_2013}%
  \BibitemOpen
  \bibfield  {author} {\bibinfo {author} {\bibfnamefont {L.}~\bibnamefont
  {Wang}}, \bibinfo {author} {\bibfnamefont {D.}~\bibnamefont {Poilblanc}},
  \bibinfo {author} {\bibfnamefont {Z.-C.}\ \bibnamefont {Gu}}, \bibinfo
  {author} {\bibfnamefont {X.-G.}\ \bibnamefont {Wen}}, \ and\ \bibinfo
  {author} {\bibfnamefont {F.}~\bibnamefont {Verstraete}},\ }\href {\doibase
  10.1103/PhysRevLett.111.037202} {\bibfield  {journal} {\bibinfo  {journal}
  {Phys. Rev. Lett.}\ }\textbf {\bibinfo {volume} {111}},\ \bibinfo {pages}
  {037202} (\bibinfo {year} {2013})}\BibitemShut {NoStop}%
\bibitem [{\citenamefont {Poilblanc}\ and\ \citenamefont
  {Mambrini}(2017)}]{Poiblanc17}%
  \BibitemOpen
  \bibfield  {author} {\bibinfo {author} {\bibfnamefont {D.}~\bibnamefont
  {Poilblanc}}\ and\ \bibinfo {author} {\bibfnamefont {M.}~\bibnamefont
  {Mambrini}},\ }\href {\doibase 10.1103/PhysRevB.96.014414} {\bibfield
  {journal} {\bibinfo  {journal} {Phys. Rev. B}\ }\textbf {\bibinfo {volume}
  {96}},\ \bibinfo {pages} {014414} (\bibinfo {year} {2017})}\BibitemShut
  {NoStop}%
\bibitem [{\citenamefont {Poilblanc}\ \emph {et~al.}(2012)\citenamefont
  {Poilblanc}, \citenamefont {Schuch}, \citenamefont {P\'erez-Garc\'{\i}a},\
  and\ \citenamefont {Cirac}}]{Poilblanc_PRB_2012}%
  \BibitemOpen
  \bibfield  {author} {\bibinfo {author} {\bibfnamefont {D.}~\bibnamefont
  {Poilblanc}}, \bibinfo {author} {\bibfnamefont {N.}~\bibnamefont {Schuch}},
  \bibinfo {author} {\bibfnamefont {D.}~\bibnamefont {P\'erez-Garc\'{\i}a}}, \
  and\ \bibinfo {author} {\bibfnamefont {J.~I.}\ \bibnamefont {Cirac}},\ }\href
  {\doibase 10.1103/PhysRevB.86.014404} {\bibfield  {journal} {\bibinfo
  {journal} {Phys. Rev. B}\ }\textbf {\bibinfo {volume} {86}},\ \bibinfo
  {pages} {014404} (\bibinfo {year} {2012})}\BibitemShut {NoStop}%
\bibitem [{\citenamefont {Chen}\ and\ \citenamefont
  {Poilblanc}(2018)}]{CJ_PRB_2018}%
  \BibitemOpen
  \bibfield  {author} {\bibinfo {author} {\bibfnamefont {J.-Y.}\ \bibnamefont
  {Chen}}\ and\ \bibinfo {author} {\bibfnamefont {D.}~\bibnamefont
  {Poilblanc}},\ }\href {\doibase 10.1103/PhysRevB.97.161107} {\bibfield
  {journal} {\bibinfo  {journal} {Phys. Rev. B}\ }\textbf {\bibinfo {volume}
  {97}},\ \bibinfo {pages} {161107} (\bibinfo {year} {2018})}\BibitemShut
  {NoStop}%
\bibitem [{\citenamefont {Poilblanc}\ \emph {et~al.}(2014)\citenamefont
  {Poilblanc}, \citenamefont {Corboz}, \citenamefont {Schuch},\ and\
  \citenamefont {Cirac}}]{Poilblanc_PRB2014}%
  \BibitemOpen
  \bibfield  {author} {\bibinfo {author} {\bibfnamefont {D.}~\bibnamefont
  {Poilblanc}}, \bibinfo {author} {\bibfnamefont {P.}~\bibnamefont {Corboz}},
  \bibinfo {author} {\bibfnamefont {N.}~\bibnamefont {Schuch}}, \ and\ \bibinfo
  {author} {\bibfnamefont {J.~I.}\ \bibnamefont {Cirac}},\ }\href {\doibase
  10.1103/PhysRevB.89.241106} {\bibfield  {journal} {\bibinfo  {journal} {Phys.
  Rev. B}\ }\textbf {\bibinfo {volume} {89}},\ \bibinfo {pages} {241106}
  (\bibinfo {year} {2014})}\BibitemShut {NoStop}%
\bibitem [{\citenamefont {Anderson}\ \emph {et~al.}(1987)\citenamefont
  {Anderson}, \citenamefont {Baskaran}, \citenamefont {Zou},\ and\
  \citenamefont {Hsu}}]{Anderson_PRL_1987}%
  \BibitemOpen
  \bibfield  {author} {\bibinfo {author} {\bibfnamefont {P.~W.}\ \bibnamefont
  {Anderson}}, \bibinfo {author} {\bibfnamefont {G.}~\bibnamefont {Baskaran}},
  \bibinfo {author} {\bibfnamefont {Z.}~\bibnamefont {Zou}}, \ and\ \bibinfo
  {author} {\bibfnamefont {T.}~\bibnamefont {Hsu}},\ }\href {\doibase
  10.1103/PhysRevLett.58.2790} {\bibfield  {journal} {\bibinfo  {journal}
  {Phys. Rev. Lett.}\ }\textbf {\bibinfo {volume} {58}},\ \bibinfo {pages}
  {2790} (\bibinfo {year} {1987})}\BibitemShut {NoStop}%
\bibitem [{\citenamefont {Kotliar}(1988)}]{Kotliar_PRB_1988}%
  \BibitemOpen
  \bibfield  {author} {\bibinfo {author} {\bibfnamefont {G.}~\bibnamefont
  {Kotliar}},\ }\href {\doibase 10.1103/PhysRevB.37.3664} {\bibfield  {journal}
  {\bibinfo  {journal} {Phys. Rev. B}\ }\textbf {\bibinfo {volume} {37}},\
  \bibinfo {pages} {3664} (\bibinfo {year} {1988})}\BibitemShut {NoStop}%
\bibitem [{\citenamefont {Anderson}(2007)}]{Anderson2007}%
  \BibitemOpen
  \bibfield  {author} {\bibinfo {author} {\bibfnamefont {P.~W.}\ \bibnamefont
  {Anderson}},\ }\href {\doibase 10.1126/science.1140970} {\bibfield  {journal}
  {\bibinfo  {journal} {Science}\ }\textbf {\bibinfo {volume} {316}},\ \bibinfo
  {pages} {1705} (\bibinfo {year} {2007})}\BibitemShut {NoStop}%
\bibitem [{\citenamefont {Verstraete}\ and\ \citenamefont
  {Cirac}(2004)}]{Verstraete04b}%
  \BibitemOpen
  \bibfield  {author} {\bibinfo {author} {\bibfnamefont {F.}~\bibnamefont
  {Verstraete}}\ and\ \bibinfo {author} {\bibfnamefont {J.~I.}\ \bibnamefont
  {Cirac}},\ }\href@noop {} {\bibfield  {journal} {\bibinfo  {journal} {eprint
  arXiv:cond-mat/0407066}\ } (\bibinfo {year} {2004})},\ \Eprint
  {http://arxiv.org/abs/arXiv:cond-mat/0407066} {arXiv:cond-mat/0407066}
  \BibitemShut {NoStop}%
\bibitem [{\citenamefont {Verstraete}\ \emph {et~al.}(2006)\citenamefont
  {Verstraete}, \citenamefont {Wolf}, \citenamefont {Perez-Garcia},\ and\
  \citenamefont {Cirac}}]{Verstraete_PRL_2006}%
  \BibitemOpen
  \bibfield  {author} {\bibinfo {author} {\bibfnamefont {F.}~\bibnamefont
  {Verstraete}}, \bibinfo {author} {\bibfnamefont {M.~M.}\ \bibnamefont
  {Wolf}}, \bibinfo {author} {\bibfnamefont {D.}~\bibnamefont {Perez-Garcia}},
  \ and\ \bibinfo {author} {\bibfnamefont {J.~I.}\ \bibnamefont {Cirac}},\
  }\href {\doibase 10.1103/PhysRevLett.96.220601} {\bibfield  {journal}
  {\bibinfo  {journal} {Phys. Rev. Lett.}\ }\textbf {\bibinfo {volume} {96}},\
  \bibinfo {pages} {220601} (\bibinfo {year} {2006})}\BibitemShut {NoStop}%
\bibitem [{\citenamefont {Jordan}\ \emph {et~al.}(2008)\citenamefont {Jordan},
  \citenamefont {Or\'us}, \citenamefont {Vidal}, \citenamefont {Verstraete},\
  and\ \citenamefont {Cirac}}]{Jordan2008}%
  \BibitemOpen
  \bibfield  {author} {\bibinfo {author} {\bibfnamefont {J.}~\bibnamefont
  {Jordan}}, \bibinfo {author} {\bibfnamefont {R.}~\bibnamefont {Or\'us}},
  \bibinfo {author} {\bibfnamefont {G.}~\bibnamefont {Vidal}}, \bibinfo
  {author} {\bibfnamefont {F.}~\bibnamefont {Verstraete}}, \ and\ \bibinfo
  {author} {\bibfnamefont {J.~I.}\ \bibnamefont {Cirac}},\ }\href {\doibase
  10.1103/PhysRevLett.101.250602} {\bibfield  {journal} {\bibinfo  {journal}
  {Phys. Rev. Lett.}\ }\textbf {\bibinfo {volume} {101}},\ \bibinfo {pages}
  {250602} (\bibinfo {year} {2008})}\BibitemShut {NoStop}%
\bibitem [{\citenamefont {Kraus}\ \emph {et~al.}(2010)\citenamefont {Kraus},
  \citenamefont {Schuch}, \citenamefont {Verstraete},\ and\ \citenamefont
  {Cirac}}]{Kraus_PRA_2010}%
  \BibitemOpen
  \bibfield  {author} {\bibinfo {author} {\bibfnamefont {C.~V.}\ \bibnamefont
  {Kraus}}, \bibinfo {author} {\bibfnamefont {N.}~\bibnamefont {Schuch}},
  \bibinfo {author} {\bibfnamefont {F.}~\bibnamefont {Verstraete}}, \ and\
  \bibinfo {author} {\bibfnamefont {J.~I.}\ \bibnamefont {Cirac}},\ }\href
  {\doibase 10.1103/PhysRevA.81.052338} {\bibfield  {journal} {\bibinfo
  {journal} {Phys. Rev. A}\ }\textbf {\bibinfo {volume} {81}},\ \bibinfo
  {pages} {052338} (\bibinfo {year} {2010})}\BibitemShut {NoStop}%
\bibitem [{\citenamefont {Corboz}\ \emph {et~al.}(2010)\citenamefont {Corboz},
  \citenamefont {Or\'us}, \citenamefont {Bauer},\ and\ \citenamefont
  {Vidal}}]{Corboz_PRB_2010}%
  \BibitemOpen
  \bibfield  {author} {\bibinfo {author} {\bibfnamefont {P.}~\bibnamefont
  {Corboz}}, \bibinfo {author} {\bibfnamefont {R.}~\bibnamefont {Or\'us}},
  \bibinfo {author} {\bibfnamefont {B.}~\bibnamefont {Bauer}}, \ and\ \bibinfo
  {author} {\bibfnamefont {G.}~\bibnamefont {Vidal}},\ }\href {\doibase
  10.1103/PhysRevB.81.165104} {\bibfield  {journal} {\bibinfo  {journal} {Phys.
  Rev. B}\ }\textbf {\bibinfo {volume} {81}},\ \bibinfo {pages} {165104}
  (\bibinfo {year} {2010})}\BibitemShut {NoStop}%
\bibitem [{\citenamefont {Bauer}\ \emph {et~al.}(2011)\citenamefont {Bauer},
  \citenamefont {Corboz}, \citenamefont {Or\'us},\ and\ \citenamefont
  {Troyer}}]{Bauer_PRB_2011}%
  \BibitemOpen
  \bibfield  {author} {\bibinfo {author} {\bibfnamefont {B.}~\bibnamefont
  {Bauer}}, \bibinfo {author} {\bibfnamefont {P.}~\bibnamefont {Corboz}},
  \bibinfo {author} {\bibfnamefont {R.}~\bibnamefont {Or\'us}}, \ and\ \bibinfo
  {author} {\bibfnamefont {M.}~\bibnamefont {Troyer}},\ }\href {\doibase
  10.1103/PhysRevB.83.125106} {\bibfield  {journal} {\bibinfo  {journal} {Phys.
  Rev. B}\ }\textbf {\bibinfo {volume} {83}},\ \bibinfo {pages} {125106}
  (\bibinfo {year} {2011})}\BibitemShut {NoStop}%
\bibitem [{\citenamefont {Singh}\ and\ \citenamefont
  {Vidal}(2012)}]{Singh_PRB2012}%
  \BibitemOpen
  \bibfield  {author} {\bibinfo {author} {\bibfnamefont {S.}~\bibnamefont
  {Singh}}\ and\ \bibinfo {author} {\bibfnamefont {G.}~\bibnamefont {Vidal}},\
  }\href {\doibase 10.1103/PhysRevB.86.195114} {\bibfield  {journal} {\bibinfo
  {journal} {Phys. Rev. B}\ }\textbf {\bibinfo {volume} {86}},\ \bibinfo
  {pages} {195114} (\bibinfo {year} {2012})}\BibitemShut {NoStop}%
\bibitem [{\citenamefont {Liu}\ \emph {et~al.}(2015)\citenamefont {Liu},
  \citenamefont {Li}, \citenamefont {Weichselbaum}, \citenamefont {von Delft},\
  and\ \citenamefont {Su}}]{Liu_PRB2015}%
  \BibitemOpen
  \bibfield  {author} {\bibinfo {author} {\bibfnamefont {T.}~\bibnamefont
  {Liu}}, \bibinfo {author} {\bibfnamefont {W.}~\bibnamefont {Li}}, \bibinfo
  {author} {\bibfnamefont {A.}~\bibnamefont {Weichselbaum}}, \bibinfo {author}
  {\bibfnamefont {J.}~\bibnamefont {von Delft}}, \ and\ \bibinfo {author}
  {\bibfnamefont {G.}~\bibnamefont {Su}},\ }\href {\doibase
  10.1103/PhysRevB.91.060403} {\bibfield  {journal} {\bibinfo  {journal} {Phys.
  Rev. B}\ }\textbf {\bibinfo {volume} {91}},\ \bibinfo {pages} {060403}
  (\bibinfo {year} {2015})}\BibitemShut {NoStop}%
\bibitem [{\citenamefont {Philipp~Schmoll}(2018)}]{Schmoll_2018}%
  \BibitemOpen
  \bibfield  {author} {\bibinfo {author} {\bibfnamefont {M.~R. R.~O.}\
  \bibnamefont {Philipp~Schmoll}, \bibfnamefont {Sukhbinder~Singh}},\ }\href
  {https://arxiv.org/abs/1809.08180} {\bibfield  {journal} {\bibinfo  {journal}
  {arXiv:1809.08180 [cond-mat.str-el]}\ } (\bibinfo {year} {2018})}\BibitemShut
  {NoStop}%
\bibitem [{\citenamefont {Hubig}(2018)}]{Hubig18}%
  \BibitemOpen
  \bibfield  {author} {\bibinfo {author} {\bibfnamefont {C.}~\bibnamefont
  {Hubig}},\ }\href {\doibase 10.21468/SciPostPhys.5.5.047} {\bibfield
  {journal} {\bibinfo  {journal} {SciPost Phys.}\ }\textbf {\bibinfo {volume}
  {5}},\ \bibinfo {pages} {47} (\bibinfo {year} {2018})}\BibitemShut {NoStop}%
\bibitem [{\citenamefont {Bruognolo}\ \emph {et~al.}(2019)\citenamefont
  {Bruognolo}, \citenamefont {Li}, \citenamefont {von Delft},\ and\
  \citenamefont {Weichselbaum}}]{Bruognolo19b}%
  \BibitemOpen
  \bibfield  {author} {\bibinfo {author} {\bibfnamefont {B.}~\bibnamefont
  {Bruognolo}}, \bibinfo {author} {\bibfnamefont {J.-W.}\ \bibnamefont {Li}},
  \bibinfo {author} {\bibfnamefont {J.}~\bibnamefont {von Delft}}, \ and\
  \bibinfo {author} {\bibfnamefont {A.}~\bibnamefont {Weichselbaum}},\
  }\href@noop {} {\bibfield  {journal} {\bibinfo  {journal} {to be published}\
  } (\bibinfo {year} {2019})}\BibitemShut {NoStop}%
\bibitem [{\citenamefont {Philipp~Schmoll}(2020)}]{Schmoll_2020}%
  \BibitemOpen
  \bibfield  {author} {\bibinfo {author} {\bibfnamefont {R.~O.}\ \bibnamefont
  {Philipp~Schmoll}},\ }\href {https://arxiv.org/abs/2005.02748} {\bibfield
  {journal} {\bibinfo  {journal} {arXiv:2005.02748 [cond-mat.str-el]}\ }
  (\bibinfo {year} {2020})}\BibitemShut {NoStop}%
\bibitem [{\citenamefont {Weichselbaum}(2012)}]{Wb12QS}%
  \BibitemOpen
  \bibfield  {author} {\bibinfo {author} {\bibfnamefont {A.}~\bibnamefont
  {Weichselbaum}},\ }\href {\doibase 10.1016/j.aop.2012.07.009} {\bibfield
  {journal} {\bibinfo  {journal} {Ann. Phys.}\ }\textbf {\bibinfo {volume}
  {327}},\ \bibinfo {pages} {2972 } (\bibinfo {year} {2012})}\BibitemShut
  {NoStop}%
\bibitem [{\citenamefont {Xie}\ \emph {et~al.}(2009)\citenamefont {Xie},
  \citenamefont {Jiang}, \citenamefont {Chen}, \citenamefont {Weng},\ and\
  \citenamefont {Xiang}}]{Xie2009}%
  \BibitemOpen
  \bibfield  {author} {\bibinfo {author} {\bibfnamefont {Z.~Y.}\ \bibnamefont
  {Xie}}, \bibinfo {author} {\bibfnamefont {H.~C.}\ \bibnamefont {Jiang}},
  \bibinfo {author} {\bibfnamefont {Q.~N.}\ \bibnamefont {Chen}}, \bibinfo
  {author} {\bibfnamefont {Z.~Y.}\ \bibnamefont {Weng}}, \ and\ \bibinfo
  {author} {\bibfnamefont {T.}~\bibnamefont {Xiang}},\ }\href {\doibase
  10.1103/PhysRevLett.103.160601} {\bibfield  {journal} {\bibinfo  {journal}
  {Phys. Rev. Lett.}\ }\textbf {\bibinfo {volume} {103}},\ \bibinfo {pages}
  {160601} (\bibinfo {year} {2009})}\BibitemShut {NoStop}%
\bibitem [{\citenamefont {Phien}\ \emph {et~al.}(2015)\citenamefont {Phien},
  \citenamefont {Bengua}, \citenamefont {Tuan}, \citenamefont {Corboz},\ and\
  \citenamefont {Or\'us}}]{Phien15}%
  \BibitemOpen
  \bibfield  {author} {\bibinfo {author} {\bibfnamefont {H.~N.}\ \bibnamefont
  {Phien}}, \bibinfo {author} {\bibfnamefont {J.~A.}\ \bibnamefont {Bengua}},
  \bibinfo {author} {\bibfnamefont {H.~D.}\ \bibnamefont {Tuan}}, \bibinfo
  {author} {\bibfnamefont {P.}~\bibnamefont {Corboz}}, \ and\ \bibinfo {author}
  {\bibfnamefont {R.}~\bibnamefont {Or\'us}},\ }\href {\doibase
  10.1103/PhysRevB.92.035142} {\bibfield  {journal} {\bibinfo  {journal} {Phys.
  Rev. B}\ }\textbf {\bibinfo {volume} {92}},\ \bibinfo {pages} {035142}
  (\bibinfo {year} {2015})}\BibitemShut {NoStop}%
\bibitem [{\citenamefont {Baxter}(1978)}]{Baxter_JSP_1978}%
  \BibitemOpen
  \bibfield  {author} {\bibinfo {author} {\bibfnamefont {R.~J.}\ \bibnamefont
  {Baxter}},\ }\href {https://doi.org/10.1007/BF01011693} {\bibfield  {journal}
  {\bibinfo  {journal} {Journal of Statistical Physics}\ }\textbf {\bibinfo
  {volume} {19}},\ \bibinfo {pages} {461} (\bibinfo {year} {1978})}\BibitemShut
  {NoStop}%
\bibitem [{\citenamefont {Nishino}\ \emph {et~al.}(1996)\citenamefont
  {Nishino}, \citenamefont {Okunishi},\ and\ \citenamefont
  {Kikuchi}}]{Nishino_PLA_1996}%
  \BibitemOpen
  \bibfield  {author} {\bibinfo {author} {\bibfnamefont {T.}~\bibnamefont
  {Nishino}}, \bibinfo {author} {\bibfnamefont {K.}~\bibnamefont {Okunishi}}, \
  and\ \bibinfo {author} {\bibfnamefont {M.}~\bibnamefont {Kikuchi}},\ }\href
  {https://doi.org/10.1016/0375-9601(96)00128-4} {\bibfield  {journal}
  {\bibinfo  {journal} {Phys. Lett. A}\ }\textbf {\bibinfo {volume} {213}},\
  \bibinfo {pages} {69} (\bibinfo {year} {1996})}\BibitemShut {NoStop}%
\bibitem [{\citenamefont {Nishino}\ and\ \citenamefont
  {Okunishi}(1996)}]{Nishino_JPJ_1996}%
  \BibitemOpen
  \bibfield  {author} {\bibinfo {author} {\bibfnamefont {T.}~\bibnamefont
  {Nishino}}\ and\ \bibinfo {author} {\bibfnamefont {K.}~\bibnamefont
  {Okunishi}},\ }\href {https://doi.org/10.1143/JPSJ.65.891} {\bibfield
  {journal} {\bibinfo  {journal} {Journal of the Physical Society of Japan}\
  }\textbf {\bibinfo {volume} {65}},\ \bibinfo {pages} {891} (\bibinfo {year}
  {1996})}\BibitemShut {NoStop}%
\bibitem [{\citenamefont {Or{\'u}s}\ and\ \citenamefont
  {Vidal}(2009)}]{Orus09}%
  \BibitemOpen
  \bibfield  {author} {\bibinfo {author} {\bibfnamefont {R.}~\bibnamefont
  {Or{\'u}s}}\ and\ \bibinfo {author} {\bibfnamefont {G.}~\bibnamefont
  {Vidal}},\ }\href {\doibase 10.1103/PhysRevB.80.094403} {\bibfield  {journal}
  {\bibinfo  {journal} {Phys. Rev. B}\ }\textbf {\bibinfo {volume} {80}},\
  \bibinfo {pages} {094403} (\bibinfo {year} {2009})}\BibitemShut {NoStop}%
\bibitem [{SI()}]{SI}%
  \BibitemOpen
  \href@noop {} {\bibinfo  {journal} {See Supplemental Material}\ }\BibitemShut
  {NoStop}%
\bibitem [{\citenamefont {Sandvik}(1997)}]{Sandvik_PRB_1997}%
  \BibitemOpen
\bibfield  {journal} {  }\bibfield  {author} {\bibinfo {author} {\bibfnamefont
  {A.~W.}\ \bibnamefont {Sandvik}},\ }\href {\doibase
  10.1103/PhysRevB.56.11678} {\bibfield  {journal} {\bibinfo  {journal} {Phys.
  Rev. B}\ }\textbf {\bibinfo {volume} {56}},\ \bibinfo {pages} {11678}
  (\bibinfo {year} {1997})}\BibitemShut {NoStop}%
\bibitem [{\citenamefont {White}\ and\ \citenamefont
  {Scalapino}(2000{\natexlab{c}})}]{white2000arxiv}%
  \BibitemOpen
  \bibfield  {author} {\bibinfo {author} {\bibfnamefont {S.~R.}\ \bibnamefont
  {White}}\ and\ \bibinfo {author} {\bibfnamefont {D.~J.}\ \bibnamefont
  {Scalapino}},\ }\href {https://arxiv.org/abs/cond-mat/0006071v1} {\bibfield
  {journal} {\bibinfo  {journal} {arXiv:cond-mat/0006071 [cond-mat.supr-con]}\
  } (\bibinfo {year} {2000}{\natexlab{c}})}\BibitemShut {NoStop}%
\bibitem [{\citenamefont {Dagotto}\ \emph {et~al.}(1992)\citenamefont
  {Dagotto}, \citenamefont {Moreo}, \citenamefont {Ortolani}, \citenamefont
  {Poilblanc},\ and\ \citenamefont {Riera}}]{Dagotto_PRB_1992}%
  \BibitemOpen
  \bibfield  {author} {\bibinfo {author} {\bibfnamefont {E.}~\bibnamefont
  {Dagotto}}, \bibinfo {author} {\bibfnamefont {A.}~\bibnamefont {Moreo}},
  \bibinfo {author} {\bibfnamefont {F.}~\bibnamefont {Ortolani}}, \bibinfo
  {author} {\bibfnamefont {D.}~\bibnamefont {Poilblanc}}, \ and\ \bibinfo
  {author} {\bibfnamefont {J.}~\bibnamefont {Riera}},\ }\href
  {https://doi.org/10.1103/PhysRevB.45.10741} {\bibfield  {journal} {\bibinfo
  {journal} {Phys. Rev. B}\ }\textbf {\bibinfo {volume} {45}},\ \bibinfo
  {pages} {10741} (\bibinfo {year} {1992})}\BibitemShut {NoStop}%
\bibitem [{\citenamefont {Dagotto}\ \emph {et~al.}(1994)\citenamefont
  {Dagotto}, \citenamefont {Riera}, \citenamefont {Chen}, \citenamefont
  {Moreo}, \citenamefont {Nazarenko}, \citenamefont {Alcaraz},\ and\
  \citenamefont {Ortolani}}]{Dagotto_PRB_1994}%
  \BibitemOpen
  \bibfield  {author} {\bibinfo {author} {\bibfnamefont {E.}~\bibnamefont
  {Dagotto}}, \bibinfo {author} {\bibfnamefont {J.}~\bibnamefont {Riera}},
  \bibinfo {author} {\bibfnamefont {Y.~C.}\ \bibnamefont {Chen}}, \bibinfo
  {author} {\bibfnamefont {A.}~\bibnamefont {Moreo}}, \bibinfo {author}
  {\bibfnamefont {A.}~\bibnamefont {Nazarenko}}, \bibinfo {author}
  {\bibfnamefont {F.}~\bibnamefont {Alcaraz}}, \ and\ \bibinfo {author}
  {\bibfnamefont {F.}~\bibnamefont {Ortolani}},\ }\href {\doibase
  10.1103/PhysRevB.49.3548} {\bibfield  {journal} {\bibinfo  {journal} {Phys.
  Rev. B}\ }\textbf {\bibinfo {volume} {49}},\ \bibinfo {pages} {3548}
  (\bibinfo {year} {1994})}\BibitemShut {NoStop}%
\bibitem [{\citenamefont {Cheng}\ \emph {et~al.}(2018)\citenamefont {Cheng},
  \citenamefont {Mondaini},\ and\ \citenamefont {Rigol}}]{CC_PRB_2018}%
  \BibitemOpen
  \bibfield  {author} {\bibinfo {author} {\bibfnamefont {C.}~\bibnamefont
  {Cheng}}, \bibinfo {author} {\bibfnamefont {R.}~\bibnamefont {Mondaini}}, \
  and\ \bibinfo {author} {\bibfnamefont {M.}~\bibnamefont {Rigol}},\ }\href
  {\doibase 10.1103/PhysRevB.98.121112} {\bibfield  {journal} {\bibinfo
  {journal} {Phys. Rev. B}\ }\textbf {\bibinfo {volume} {98}},\ \bibinfo
  {pages} {121112} (\bibinfo {year} {2018})}\BibitemShut {NoStop}%
\bibitem [{\citenamefont {Tranquada}\ \emph {et~al.}(1995)\citenamefont
  {Tranquada}, \citenamefont {Sternlieb}, \citenamefont {Axe}, \citenamefont
  {Nakamura},\ and\ \citenamefont {Uchida}}]{Tranquada_NATURE_1995}%
  \BibitemOpen
  \bibfield  {author} {\bibinfo {author} {\bibfnamefont {J.~M.}\ \bibnamefont
  {Tranquada}}, \bibinfo {author} {\bibfnamefont {B.~J.}\ \bibnamefont
  {Sternlieb}}, \bibinfo {author} {\bibfnamefont {J.~D.}\ \bibnamefont {Axe}},
  \bibinfo {author} {\bibfnamefont {Y.}~\bibnamefont {Nakamura}}, \ and\
  \bibinfo {author} {\bibfnamefont {S.}~\bibnamefont {Uchida}},\ }\href
  {https://doi.org/10.1038/375561a0} {\bibfield  {journal} {\bibinfo  {journal}
  {Nature}\ }\textbf {\bibinfo {volume} {375}},\ \bibinfo {pages} {561}
  (\bibinfo {year} {1995})}\BibitemShut {NoStop}%
\bibitem [{\citenamefont {Kivelson}\ \emph {et~al.}(2003)\citenamefont
  {Kivelson}, \citenamefont {Bindloss}, \citenamefont {Fradkin}, \citenamefont
  {Oganesyan}, \citenamefont {Tranquada}, \citenamefont {Kapitulnik},\ and\
  \citenamefont {Howald}}]{Kivelson_RMP_2003}%
  \BibitemOpen
  \bibfield  {author} {\bibinfo {author} {\bibfnamefont {S.~A.}\ \bibnamefont
  {Kivelson}}, \bibinfo {author} {\bibfnamefont {I.~P.}\ \bibnamefont
  {Bindloss}}, \bibinfo {author} {\bibfnamefont {E.}~\bibnamefont {Fradkin}},
  \bibinfo {author} {\bibfnamefont {V.}~\bibnamefont {Oganesyan}}, \bibinfo
  {author} {\bibfnamefont {J.~M.}\ \bibnamefont {Tranquada}}, \bibinfo {author}
  {\bibfnamefont {A.}~\bibnamefont {Kapitulnik}}, \ and\ \bibinfo {author}
  {\bibfnamefont {C.}~\bibnamefont {Howald}},\ }\href
  {https://doi.org/10.1103/RevModPhys.75.1201} {\bibfield  {journal} {\bibinfo
  {journal} {Rev. Mod. Phys.}\ }\textbf {\bibinfo {volume} {75}},\ \bibinfo
  {pages} {1201} (\bibinfo {year} {2003})}\BibitemShut {NoStop}%
\bibitem [{\citenamefont {Vojta}(2009)}]{Vojta_2009AIP}%
  \BibitemOpen
  \bibfield  {author} {\bibinfo {author} {\bibfnamefont {M.}~\bibnamefont
  {Vojta}},\ }\href {\doibase 10.1080/00018730903122242} {\bibfield  {journal}
  {\bibinfo  {journal} {Advances in Physics}\ }\textbf {\bibinfo {volume}
  {58}},\ \bibinfo {pages} {699} (\bibinfo {year} {2009})}\BibitemShut
  {NoStop}%
\bibitem [{\citenamefont {Robinson}\ \emph {et~al.}()\citenamefont {Robinson},
  \citenamefont {Johnson}, \citenamefont {Rice},\ and\ \citenamefont
  {Tsvelik}}]{Robinson2019}%
  \BibitemOpen
  \bibfield  {author} {\bibinfo {author} {\bibfnamefont {N.~J.}\ \bibnamefont
  {Robinson}}, \bibinfo {author} {\bibfnamefont {P.~D.}\ \bibnamefont
  {Johnson}}, \bibinfo {author} {\bibfnamefont {T.~M.}\ \bibnamefont {Rice}}, \
  and\ \bibinfo {author} {\bibfnamefont {A.~M.}\ \bibnamefont {Tsvelik}},\
  }\href {http://arxiv.org/abs/1906.09005v1} {\bibinfo  {journal}
  {arXiv:1906.09005}\ }\BibitemShut {NoStop}%
\end{thebibliography}
